\documentclass{article}\usepackage[]{graphicx}\usepackage[]{xcolor}
\makeatletter
\def\maxwidth{ %
  \ifdim\Gin@nat@width>\linewidth
    \linewidth
  \else
    \Gin@nat@width
  \fi
}
\makeatother

\definecolor{fgcolor}{rgb}{0.345, 0.345, 0.345}

\usepackage{framed}
\makeatletter
 {\par\unskip\endMakeFramed%
 \at@end@of@kframe}
\makeatother

\definecolor{shadecolor}{rgb}{.97, .97, .97}
\definecolor{messagecolor}{rgb}{0, 0, 0}
\definecolor{warningcolor}{rgb}{1, 0, 1}
\definecolor{errorcolor}{rgb}{1, 0, 0}
\newenvironment{knitrout}{}{} 

\usepackage{alltt}

\usepackage{amsmath}
\usepackage{amssymb}
\usepackage{amsthm}
\usepackage[left=1in,right=1in,top=1in,bottom=1in]{geometry}
\usepackage{graphicx}
\usepackage{url}
\usepackage{authblk}
\usepackage{enumitem}
\usepackage{multirow}
\usepackage{booktabs}
\usepackage{caption}
\usepackage{subcaption}
\usepackage[section]{placeins} 
\usepackage{algorithm}
\usepackage{algpseudocode}
\usepackage{xr}
\usepackage[numbers]{natbib}

\theoremstyle{plain} 
\newtheorem{assum}{Assumption}
\newtheorem{prop}{Proposition}

\newtheorem{corollary}{Corollary}

\theoremstyle{definition} 
\newtheorem{defn}{Definition}
\newtheorem{ex}{Example}

\newcommand{\R}{\mathbb{R}}
\newcommand{\Prob}{\mathsf{P}}

\newcommand{\E}{\mathbb{E}\,}

\newcommand{\J}{\mathcal{J}}
\newcommand{\proxy}{\mathcal{R}}
\newcommand{\N}{\mathcal{N}}

\newcommand{\ml}[1]{\mathit{#1}} 
\newcommand{\cl}{\,{:}\,}

\newcommand*\di{\mathop{}\!\mathrm{d}} 
\newcommand{\MLE}{\ml{MLE}}
\newcommand{\MESLE}{\ml{MESLE}}

\newcommand{\Var}{\text{Var}}
\newcommand{\vech}{\text{vech}}
\newcommand{\dimpar}{d}
\newcommand{\matr}{\text{mat}}

\newcommand{\cqc}{P} 
\newcommand{\proj}{S} 

\usepackage{pdfpages}

\title{Scalable simulation-based inference for implicitly defined models using a metamodel for Monte Carlo log-likelihood estimator}
\author{Joonha Park}
\affil{Department of Mathematics, University of Kansas, Lawrence, KS 66045 USA \\(email: {j.park@ku.edu})}
\date{}

\IfFileExists{upquote.sty}{\usepackage{upquote}}{}
\begin{document}
\maketitle

\begin{abstract}
  Models implicitly defined through a random simulator of a process have become widely used in scientific and industrial applications in recent years.
  However, simulation-based inference methods for such implicit models, like approximate Bayesian computation (ABC), often scale poorly as data size increases.
  We develop a scalable inference method for implicitly defined models using a metamodel for the Monte Carlo log-likelihood estimator derived from simulations.
  This metamodel characterizes both statistical and simulation-based randomness in the distribution of the log-likelihood estimator across different parameter values.
  Our metamodel-based method quantifies uncertainty in parameter estimation in a principled manner, leveraging the local asymptotic normality of the mean function of the log-likelihood estimator.
  We apply this method to construct accurate confidence intervals for parameters of partially observed Markov process models where the Monte Carlo log-likelihood estimator is obtained using the bootstrap particle filter.
  We numerically demonstrate that our method enables accurate and highly scalable parameter inference across several examples, including a mechanistic compartment model for infectious diseases.
\end{abstract}

\textbf{keywords:}
parameter inference, simulation-based inference, implicitly defined model, Monte Carlo methods, particle filter


\section{Introduction}\label{sec:intro}

A model is said to be \emph{implicitly defined} if it is defined using a random simulator of the underlying process \cite{diggle1984monte}.
If the density function of an implicitly defined process cannot be evaluated, traditional likelihood-based parameter inference is generally impossible, because the likelihood function is not analytically tractable.
In this paper, we develop a scalable, simulation-based parameter inference method for partially observed, implicitly defined models.
Simulation models are widely used for a variety of industrial and scientific applications, precipitated by recent advances in the digital twin technology \cite{laubenbacher2021using, subramanian2021quantifying, tao2019make, national2024foundational}.

\begin{figure}[t]
\begin{knitrout}
\definecolor{shadecolor}{rgb}{0.969, 0.969, 0.969}\color{fgcolor}

{\centering \includegraphics[width=\maxwidth]{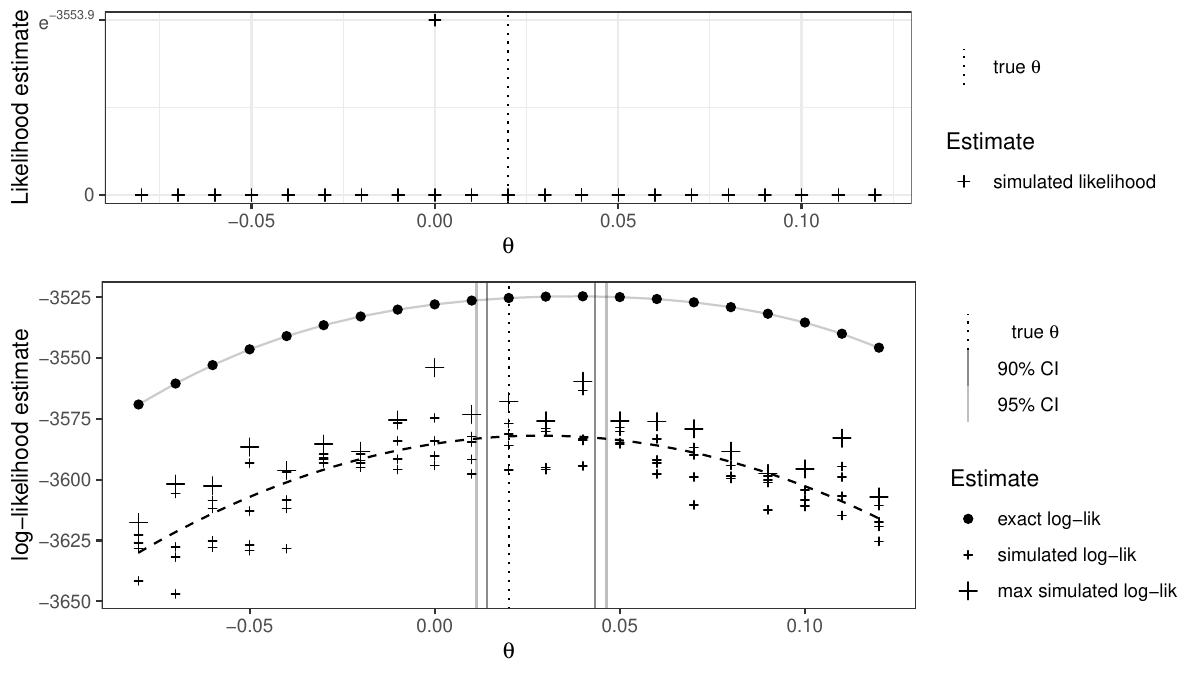} 

}

\end{knitrout}
\caption{
  The top panel shows the likelihood estimates obtained by running the particle filter five times for each $\theta$ value on a natural (non-logarithmic) scale.
  The bottom panel shows the log of those likelihood estimates, marked by `+' symbols.
  The maximum among the five values for each $\theta$ is indicated by a larger `+' symbol.
  The dashed curve shows the fitted quadratic polynomial.
  The exact log-likelihoods computed using the Kalman filter are indicated by `$\bullet$'.
  The dotted vertical line indicates the true parameter value, $\theta=0.02$, and the gray vertical lines indicate the constructed 90\% and 95\% confidence intervals.
}
\label{fig:intro_OU}
\end{figure}

We consider a collection of observations $Y_{1:n}$ that are conditionally independent given a realization $X$ of the random process.
For example, $Y_{1:n}$ may be noisy or partial observations of a Markov process $X_{1:n}$, where $Y_i$ are conditionally independent observation of $X_i$, $i\in 1\cl n$.
Generally, we assume that the implicitly defined process $X$ is parameterized by $\theta\in\mathsf\Theta$.
If we denote the measurement density of the observed data $y=y_{1:n}$ given $X=x$ by $g(y|x;\theta)$, the likelihood is given by
\begin{equation}
  L(\theta;y) = \int g(y|x;\theta) \di P_\theta(x).
  \label{eqn:lik}
\end{equation}
Here $\theta$ parametrizes both the law of the latent process and the measurement process, but it may consist of two separate components each governing only one of the processes.
In principle, the likelihood \eqref{eqn:lik} may be estimated by
\begin{equation}
  \hat L(\theta;y) = \frac{1}{J} \sum_{j=1}^J g(y|X_j(\theta);\theta)
  \label{eqn:lik_est_unbiased}
\end{equation}
where $X_j(\theta)$, $j\in 1\cl J$, are independent simulations of $P_\theta$.
Although the Monte Carlo likelihood estimator \eqref{eqn:lik_est_unbiased} is unbiased for $L(\theta;y)$, its variance grows exponentially with an increasing size of the data $y_{1:n}$.
For partially observed Markov process (POMP) models, an unbiased estimator of the likelihood with substantially smaller variance than that of \eqref{eqn:lik_est_unbiased} can be obtained by running the particle filter \cite{cappe2007overview, delmoral2004feynman, doucet2001introduction}.
In particular, when the latent Markov process is implicitly defined via a simulator, an unbiased Monte Carlo likelihood estimator can be obtained using the bootstrap particle filter \cite{gordon1993novel, he2009plug}.
However, the variance of this estimator still scales exponentially with increasing $n$ \cite{cerou2011nonasymptotic}.
The exponentially large Monte Carlo variance in the likelihood estimator is typically realized by a very small probability of attaining relatively extremely large values.
The highly skewed distribution of the likelihood estimator implies that inference based on $\hat L(\theta;y)$ faces a great amount of Monte Carlo variability. 
This phenomenon is illustrated by the top panel of Figure~\ref{fig:intro_OU}, which displays the likelihood estimates independently obtained five times for each $\theta$ value for a certain POMP model given by Example~\ref{ex:intro} below.

In order to enable simulation-based inference that is scalable to large data, methods that involve approximation of the \emph{log}-likelihood function have been developed.
These methods utilize the fact that the logarithm of Monte Carlo likelihood estimators often have distributions with manageable variance and skewness.
Diggle and Gratton \cite{diggle1984monte} considered parameter estimation where independent and identically distributed (iid) draws from an implicit model $X_i \sim P_\theta$, $i\in 1\cl n$, are observed without noise.
They proposed a method where the log-likelihood function estimated using simulations and kernel density estimation is maximized by a stochastic optimization procedure such as the Nelder-Mead algorithm \cite{nelder1965simplex}.
However, Diggle and Gratton \cite{diggle1984monte} only provided a heuristic for estimating the parameter uncertainty, for situations where the Monte Carlo error is negligibly small compared to the statistical error.
Such precise estimation of the log-likelihood function using a kernel method may be possible when the observations are iid and low-dimensional, but it is generally impossible for dependent or moderately high dimensional observations.

Ionides et al.\@ \cite{ionides2017monte} developed a simulation-based inference method for partially observed, implicit models producing dependent observations.
They considered estimating the log-likelihood function using the log of the Monte Carlo likelihood estimates such as those obtained by the bootstrap particle filter. 
Both Ionides et al.\@ \cite{ionides2017monte} and the current paper consider situations where the random error in the Monte Carlo log-likelihood estimator is not small compared to the statistical error.
Ionides et al.\@ \cite{ionides2017monte} used a metamodel that locally approximates the log-likelihood function with a quadratic polynomial, justified by the local asymptotic normality (LAN) of the log-likelihood function \cite{lecam2000asymptotics, vandervaart1998asymptotic}.
Local or global metamodels are widely used for simulation-based optimization in response surface methodology \cite{barton2006metamodel}.
Ionides et al.\@ \cite{ionides2017monte} constructed profile confidence intervals by approximating the uncertainty in parameter estimation by the sum of the Monte Carlo variance due to random simulations and the statistical variance derived from a standard asymptotic theory for maximum likelihood estimation.
The Monte Carlo variance was approximately estimated by a quadratic regression and the use of the delta method. 
However, Ionides et al.\@ \cite{ionides2017monte} overlooked the fact that, in general, the log-likelihood function $\ell(\theta)$ and the mean function of the Monte Carlo log-likelihood estimator $\mu(\theta)$ have different curvatures and distinct statistical properties due to data randomness.
This can result in misquantified uncertainty, further increasing the inference bias beyond that introduced by the delta method approximation.

In the current paper, we refine the metamodel-based uncertainty quantification approach by Ionides et al.\@ \cite{ionides2017monte}.
We develop a local asymptotic normality (LAN) theory for the mean function $\mu(\theta;Y)$ of the log-likelihood estimator.
This LAN property forms a basis for our metamodel describing the statistical properties of $\mu(\theta;Y)$.
Our use of a carefully characterized metamodel reduces the biases in the method by Ionides et al.\@ \cite{ionides2017monte}, which incorrectly relies on the observed Fisher information to quantify statistical error.
In contrast, we use the observed data to quantify statistical uncertainty, assuming that the observations $y_{1:n}$ can be divided into subsets that are nearly independent of each other.
This assumption is often satisfied by stochastic processes with reasonable mixing properties.

We illustrate our method through the following example.


\begin{ex}\label{ex:intro}
  Consider a multivariate autoregressive process with partial observations $Y_i$:
  \begin{equation*}
    X_i = A X_{i-1} + v_i, \qquad
    Y_i = X_i + e_i, \qquad i=1\cl 200.
    \label{eqn:intro_OU}
  \end{equation*}
  Here $X_i, Y_i \in \mathbb R^{10}$, $A\in \mathbb R^{10\times 10}$, $v_i \overset{iid}{\sim} \N(0, I_{10})$, and $e_i \overset{iid}{\sim} \N(0, I_{10})$.
  The matrix $A$ has diagonal entries all equal to -0.3 and off-diagonal entries all equal to $\theta$.
  We generate an observation sequence $y_{1:200}$ at $\theta = 0.02$.
  The bootstrap particle filter is run with two hundred particles five times for each parameter value $\theta \in \{-0.08, -0.07, \dots, 0.12\}$.
  We will refer to an estimate of the log-likelihood that is obtained through simulations as a \emph{simulated log-likelihood}.
\end{ex}

  Figure~\ref{fig:intro_OU} shows five simulated log-likelihoods for each $\theta$, denoted by $\ell^S_k(\theta)$, $k \in 1\cl 5$.
  For comparison, the exact log-likelihood for each $\theta$ is shown, obtained by running the Kalman filter \cite{kalman1960new}.
  

  We employ a metamodel in which simulated log-likelihoods are assumed to be approximately normally distributed with a locally quadratic mean function $\mu(\theta; y_{1:n})$.
  The function $\mu(\theta; y_{1:n})$ is estimated by fitting a quadratic polynomial through the simulated log-likelihoods.
  The statistical properties of $\mu(\theta; Y_{1:n})$ where the observations $Y_{1:n}$ are generated under a certain parameter value are characterized by the local asymptotic normality (LAN) of $\mu(\theta; Y_{1:n})$ (see Section~\ref{sec:sbi_lan} for details.)
  A variance parameter in the LAN statement for $\mu(\theta; Y_{1:n})$ is estimated by assuming approximate independence of contiguous subsets of $Y_{1:n}$ (Section~\ref{sec:K1est}.)
  Confidence intervals are constructed using the metamodel for the random function $\mu(\theta; Y_{1:n})$, incorporating both sources of randomness originating from the observations and the simulations (Section~\ref{sec:est_proxy}.) 
Our theory and method are developed generally for multiple parameters, unlike the approach proposed by Ionides et al.\@ \cite{ionides2017monte}, which focuses on one-dimensional profile likelihoods.

Relative to other methods for simulation-based inference, our approach is highly scalable with increasing data size.
Cranmer et al.\@ \cite{cranmer2020frontier} provide a survey of traditional and recent simulation-based inference algorithms.
Approximate Bayesian computation (ABC) estimates the likelihood using the proportion of simulations in close proximity of the observed data \cite{beaumont2002approximate}.
Traditional inference methods such as ABC experience the curse of dimensionality similar to that affecting kernel density estimation in high dimensional spaces.
This issue is often addressed by using summary statistics, at the expense of introducing a bias and a loss of efficiency due to the information reduction \cite{sisson2018handbook}.
If partial observations of the implicitly defined process are available and the density of the observations given the latent states is analytically tractable, pseudo-marginal Markov chain Monte Carlo (MCMC) methods can enable exact Bayesian inference by utilizing unbiased Monte Carlo likelihood estimators \cite{beaumont2003estimation, andrieu2009pseudo, andrieu2010particle}.
However, these methods are also affected by the curse of dimensionality, which is manifested by an exponentially scaling Monte Carlo variance and thus a poorly scaling sampling efficiency \cite{sherlock2015efficiency, doucet2015efficient}.

Recent advances in machine learning have enabled training a surrogate model for the distribution of observations $Y$ through neural density estimation techniques such as normalizing flows \cite{dinh2014nice, dinh2017density, papamakarios2017masked}.
We note that these surrogate models, which approximate the distribution of $Y$ for each given $\theta$, differ from the metamodels we consider, which describe the log-likelihood $\ell(\theta;y)$ as a function of $\theta$ for given $y$.
These machine learning techniques have the advantage that, once a surrogate model is trained using simulations, inference using the approximate likelihoods is straightforward.
However, training a high-fidelity surrogate model for a high dimensional observation distribution requires a large number of simulations and may still involve substantial error in some parts of the observation space.
Moreover, quantifying the uncertainty associated with the trained surrogate model is often challenging.

Our metamodel-based inference approach offers several advantages compared to these machine-learning-based methods.
First, since Monte Carlo log-likelihood estimators often have distributions that scale linearly rather than exponentially with the size of the data $y=y_{1:n}$, our method is scalable to models that produce large amounts of data.
Second, it enjoys a favorable sampling efficiency because the metamodel enables pooling the information across $\theta$ values.
Third, it offers a principled method for uncertainty quantification, since the characterization of the variation in $\mu(\theta;Y)$ by the LAN property is consistent.
We note that recent machine learning techniques may be combined with the metamodel-based inference framework we propose in this work by regarding the logarithm of the approximate likelihood derived from the trained surrogate model as a Monte Carlo log-likelihood estimator.
However, we do not delve into this in the current paper. 

The rest of the paper is organized as follows.
Section~\ref{sec:sbi} lays a theoretical basis for our scalable simulation-based inference approach.
We develop a local asymptotic normality (LAN) property for $\mu(\theta;Y)$ and set up a simulation metamodel.
Section~\ref{sec:est_mesle_K1} concerns fitting the metamodel to the simulated log-likelihoods via quadratic regression.
Section~\ref{sec:est_proxy} develops a procedure for uncertainty quantification and construction of confidence intervals, taking into account both simulation randomness and observation randomness.
Section~\ref{sec:numerical} provides numerical results showing the accuracy of our simulation-based inference method using several examples, including a mechanistic model for the population dynamics of an infectious disease.
In Section~\ref{sec:comp_pmcmc}, we numerically compare the scalability of our method and that of a pseudo-marginal MCMC method.
Section~\ref{sec:autotuning} introduces two automatic tuning algorithms that enhance the applicability of our method.
One algorithm automatically adjusts the weights associated with simulations to control the bias introduced by the quadratic approximation while maintaining high estimation efficiency.
The other algorithm identifies approximately optimal points for subsequent simulations.
Section~\ref{sec:discussion} concludes with discussion.

\section{Simulation-based inference}\label{sec:sbi}
In this section, we define a simulation metamodel and relevant quantities.
Section~\ref{sec:sim_log_lik} outlines the difference between traditional likelihood-based inference and our metamodel-based inference.
Inference using a metamodel involves estimation of what we call a \emph{simulation-based proxy}, which is the maximizer of the mean function of the log-likelihood estimator averaged over the data distribution.
Section~\ref{sec:sbi_lan} develops the local asymptotic normality (LAN) for the mean function of the log-likelihood estimator.
Section~\ref{sec:metamodel} defines a metamodel for the log-likelihood estimator, based on the LAN property.
Section~\ref{sec:sbi_pomp} discusses application to partially observed Markov process (POMP) models in the case where the simulation-based log-likelihood estimator is obtained using the bootstrap particle filter.
Section~\ref{sec:bias} discusses the bias introduced by a simulation metamodel.

\subsection{Simulation-based proxy}\label{sec:sim_log_lik}
We denote the marginal density of the observation $Y = y$ under parameter $\theta$ by
\[
  p^Y_\theta(y) = \frac{\di P^Y_\theta}{\di y} = \int g(y|x;\theta) \di P_\theta(x).
\]
If we simulate the underlying process and obtain a draw $X$ from $P_\theta$, we may consider $\log g(y|X;\theta)$ as an estimate of the log-likelihood $\ell(\theta; y) = \log p^Y_\theta(y)$.
We will refer to an estimate of the log-likelihood obtained via simulation of the underlying process as a \emph{simulated log-likelihood}.
The simulated log-likelihood will be denoted by $\ell^S(\theta;y)$ or $\ell^S(\theta)$, or in some cases by $\ell^S(X,y)$ when emphasizing the fact that it is a function of the simulated draw $X$.
We define the \emph{expected simulated log-likelihood} as
\begin{equation*}
  \mu(\theta; y) := \E \ell^S(\theta;y) = \int \ell^S(x,y) \di P_\theta(x).
  \label{eqn:esll}
\end{equation*}
If the simulated log-likelihood $\ell^S(\theta;y)$ is given by the logarithm of an unbiased likelihood estimator $\hat L(\theta;y)$, such as $g(y|X;\theta)$ where $X$ is a draw from $P_\theta$, then $\mu(\theta;y)$ is less than or equal to the exact log-likelihood $\ell(\theta;y)$ due to Jensen's inequality:
\begin{equation*}
  \mu(\theta;y)
  =\E \ell^S(\theta;y)
  \leq \log \E \hat L(\theta; y)
  = \ell(\theta;y).
  \label{eqn:jensen}
\end{equation*}
The Jensen bias can be written as
\begin{equation}
  \ell(\theta; y) - \mu(\theta; y) = \sum_{j\geq 2} \frac{\kappa_j(\theta; y)}{j!}
  \label{eqn:bias_expansion}
\end{equation}
where $\kappa_j(\theta;y)$ is the $j$-th order cumulant of the distribution of the simulated log-likelihood $\ell^S(\theta;y)$ for given $y$ (see the supplementary text, Section~\ref{sec:Jensenbias} for more details.)
Since the Jensen bias depends on higher-order cumulants of $\ell^S(\theta;y)$, or the tail distribution of $\ell^S(\theta;y)$, estimating this bias involves high Monte Carlo variability. 

For a given set of observations $y = y_{1:n}$, the maximizer of the expected simulated log-likelihood $\mu(\theta;y)$ will be referred to as the \emph{maximum expected simulated log-likelihood estimator} (\emph{MESLE}).
\begin{defn}[MESLE]\label{defn:mesle}
  Given the observations $y_{1:n}$, the maximum expected simulated log-likelihood estimate (MESLE) is defined as 
  \begin{equation}
    \theta_\MESLE(y_{1:n}) := \arg\max_\theta \mu(\theta;y_{1:n}).
    \label{eqn:mesle}
  \end{equation}
\end{defn}
The MESLE substitutes the role of the the maximum likelihood estimator (MLE) for traditional likelihood-based inference.
Since $\mu(\theta;y)$ is not known analytically and is instead estimated through the simulated log-likelihoods, the MESLE must also be estimated via simulations.
If the Jensen bias $\ell(\theta;y) - \mu(\theta;y)$ varies as a function of $\theta$, the MESLE may differ from the MLE.

If the observed data $Y$ are generated under $\theta_0$, the \emph{data-averaged expected simulated log-likelihood} is defined as
\begin{equation}
  U(\theta_0, \theta) := \E_{\theta_0} \mu(\theta; Y) = \int \mu(\theta;y) p^Y_{\theta_0}(y) \di y.
  \label{eqn:daesll}
\end{equation}
The parameter value $\theta$ that maximizes $U(\theta_0,\theta)$ will be referred to as the \emph{simulation-based proxy} for $\theta_0$.
\begin{defn}[Simulation-based proxy]\label{defn:proxy}
  The \emph{simulation-based proxy for $\theta_0$}, denoted by $\proxy(\theta_0)$, is defined as the maximizer of the data-averaged expected simulated log-likelihood $U(\theta_0,\theta)$:
  \[
    \proxy(\theta_0) = \underset{\theta}{\arg\max}\, U(\theta_0, \theta).
  \]
For simplicity, $\proxy(\theta_0)$ will sometimes be denoted by $\theta_*$.
\end{defn}
The MESLE converges in probability to the simulation-based proxy as the data size increases, under certain regularity conditions for M-estimation \cite{vandervaart1998asymptotic}.

If the Jensen bias $\ell(\theta;Y_{1:n}) - \mu(\theta;Y_{1:n})$ is constant as a function of $\theta$ for every $Y_{1:n}$, then $U(\theta_0, \theta)$ will have a constant difference with respect to $\E_{\theta_0} \ell(\theta; Y)$, which is maximized at $\theta = \theta_0$.
Thus the inference bias, $\Vert \proxy(\theta_0) - \theta_0 \Vert$, will be equal to zero.
However, $\proxy(\theta_0)$ can be different from $\theta_0$ in general (see supplementary Section~\ref{sec:supp_ex_sim_log_lik} for concrete examples.)
Any simulation-based inference method that estimates a response surface for $\ell(\theta)$ using a metamodel inherently involves the possibility of nonzero inference bias, although the method may be scalable.
In contrast, methods like ABC or pseudo-marginal MCMC, which estimate the likelihood $L(\theta)$ separately for each $\theta$ without relying on a metamodel, may be asymptotically exact as the number of simulation increases but are not scalable.
When the data size is large, the latter class of methods is unlikely to yield smaller inference errors than the methods that use a metamodel.
A numerical comparison between our method and a pseudo-marginal MCMC method, presented in Section~\ref{sec:comp_pmcmc}, supports this.

\subsection{Local asymptotic normality for $\mu(\theta;y_{1;n})$}\label{sec:sbi_lan}

We develop the local asymptotic normality (LAN) for the mean function $\mu(\theta;y_{1:n})$ of the simulated log-likelihood $\ell^S(\theta;y_{1:n})$ under several assumptions.
First, we assume that $\ell^S(\theta;y_{1:n})$ can be expressed as the sum of log-likelihood estimates for individual observation pieces $y_i$, $i\in 1\cl n$.

\begin{assum}\label{assum:simll_sum}
  A log-likelihood estimator $\ell^S_i(\theta; y_i)$ is available for each observation $y_i$, $i \in 1\cl n$.
  These individual simulated log-likelihoods sum to the simulated log-likelihood for the entire data $y_{1:n}$:
  \[
    \ell^S(\theta;y_{1:n}) = \sum_{i=1}^n \ell^S_i(\theta;y_i).
  \]
\end{assum}

For example, Assumption~\ref{assum:simll_sum} is satisfied if $y_i$, $i\in 1\cl n$, are conditionally independent given $X$ with measurement density $g(y_{1:n}|X) = \prod_{i=1}^n g_i(y_i|X)$ and if the simulated log-likelihood is given by
\[
  \ell^S(\theta; y_{1:n}) = \log g(y_{1:n}|X) = \sum_{i=1}^n \log g_i(y_i|X),  \quad X \sim P_\theta,
\]
where we let $\ell^S_i(\theta; y_i) = \log g_i(y_i|X)$.
Assumption~\ref{assum:simll_sum} is also satisfied when the simulated log-likelihood is obtained through the bootstrap particle filter for POMP models.
See Section~\ref{sec:sbi_pomp} for details.
Under Assumption~\ref{assum:simll_sum}, we let $\mu_i(\theta;y_i) = \E \ell_i^S(\theta; y_i)$ such that expected simulated log-likelihood can be expressed as
\[
  \mu(\theta; y_{1:n})
  = \sum_{i=1}^n \mu_i(\theta; y_i).
\]

We make the following additional assumptions.
Recall that when the true parameter is $\theta_0$, the simulation-based proxy is denoted by $\proxy(\theta_0) = \theta_*$.
\begin{assum}\label{assum:C3mu}
  The parameter space $\mathsf \Theta$ is a subset of $\R^d$, and the individual expected simulated log-likelihood $\mu_i(\theta; y_i)$ is three times continuously differentiable with respect to $\theta = (\theta_{(1)}, \dots, \theta_{(d)})$ for every $y_i$, $i\in 1\cl n$.
\end{assum}
\begin{assum}\label{assum:1st_diff}
  There exists a positive definite matrix $K_1(\theta_0) \in \R^{d \times d}$ such that as $n\to\infty$,
  \[
    \frac{1}{\sqrt n} \sum_{i=1}^n \frac{\partial \mu_i}{\partial \theta}(\theta_*; Y_i) \underset{Y_{1:n}\sim P^Y_{\theta_0}}{\Longrightarrow} \N(0, K_1(\theta_0)),
  \]
  where $\Rightarrow$ indicates convergence in distribution.
\end{assum}
\begin{assum}\label{assum:2nd_diff}
  There exists a positive definite matrix $K_2(\theta_0) \in \R^{d \times d}$ such that
  \[
    \frac{1}{n} \sum_{i=1}^n \frac{\partial^2 \mu_i}{\partial \theta^2} (\theta_*; Y_i) \overset{i.p.}{\underset{Y_{1:n}\sim P^Y_{\theta_0}}{\longrightarrow}} -K_2(\theta_0),
  \]
  where $\overset{i.p.}{\to}$ indicates convergence in probability.
\end{assum}
\begin{assum}\label{assum:3rd_diff}
  There exists an open ball $B_0$ containing $\theta_*$ and a constant $C$ such that for every $i$,
  \[
    \E \sup_{\theta\in B_0} \max_{k_1,k_2,k_3 \in 1:d} \left\Vert \frac{\partial^3 \mu_i}{\partial \theta_{(k_1)}\partial \theta_{(k_2)} \partial \theta_{(k_3)}} (\theta; Y_i) \right\Vert \leq C.
  \]
\end{assum}

Assumptions~\ref{assum:1st_diff} and \ref{assum:2nd_diff} are satisfied under mild conditions when $\ell^S_i(\theta; Y_i)$ are marginally independent, due to the central limit theorem and the law of large numbers, respectively.
The observations $Y_{1:n}$ are marginally independent, for example, when the components of $X=(X_1,\dots, X_n)$ are independent and $Y_i$ only depends on $X_i$, for $i\in 1\cl n$.
However, Assumptions~\ref{assum:1st_diff} and \ref{assum:2nd_diff} may still be satisfied for marginally dependent $Y_i$ if $X$ satisfies certain mixing conditions, as discussed in Supplementary Section~\ref{sec:supp_mixing}.

Under Assumptions~\ref{assum:simll_sum}-\ref{assum:3rd_diff}, we obtain the following LAN result for $\mu(\theta;Y_{1:n})$.
\begin{prop}[Local asymptotic normality (LAN) for $\mu(\theta;Y_{1:n})$]\label{prop:lan_sim}
  Suppose that Assumptions~\ref{assum:simll_sum}-\ref{assum:3rd_diff} hold where $Y_{1:n} \sim P^Y_{\theta_0}$.
  Let
  \[
    S_n := \frac{1}{\sqrt n} \sum_{i=1}^n \frac{\partial \mu_i}{\partial \theta}(\theta_*; Y_i).
  \]
  Then $S_n$ converges in distribution to $\N(0, K_1(\theta_0))$ and  
  \begin{equation}
    \mu(\theta_* + \tfrac{t}{\sqrt n}; Y_{1:n}) - \mu(\theta_*; Y_{1:n})
    - S_n^\top t + \frac{1}{2} t^\top K_2(\theta_0) t
    \label{eqn:lan}
  \end{equation}
  converges in probability to zero uniformly for $t\in B$ as $n\to\infty$ for every bounded set $B$ containing 0. 
\end{prop}

The proof of Proposition~\ref{prop:lan_sim} is given in the supplementary text, Section~\ref{sec:proofs_sbi_lan}.
Proposition~\ref{prop:lan_sim} can be compared to the original LAN for the log-likelihood function $\ell(\theta)$, which states that there exists a sequence of random variables $\{S_n'; n\geq 1\}$ that converges in probability to $\N(0, \mathcal I(\theta_0))$ and such that
\[
  \ell(\theta_0 + \tfrac{t}{\sqrt n}; Y_{1:n}) - \ell(\theta_0; Y_{1:n})
  = S_n'^\top t - \frac{1}{2} t^\top \mathcal I(\theta_0) t + o_p(1)
\]
as $n\to\infty$ for every $t$ \cite{lecam2000asymptotics}.
Iid observations $Y_{1:n}$ satisfy this LAN property under certain regularity conditions with $\mathcal I(\theta_0)$ equal to the Fisher information \cite{vandervaart1998asymptotic}. 
We summarize the differences between the original LAN and Proposition~\ref{prop:lan_sim} as follows:
\begin{enumerate}
  \setlength\itemsep{0ex}
\item Proposition~\ref{prop:lan_sim} suggests an asymptotically quadratic approximation for the expected simulated log-likelihood $\mu(\theta; Y_{1:n})$, not for the log-likelihood function $\ell(\theta; Y_{1:n})$.
\item Whereas the original LAN for $\ell(\theta; Y_{1:n})$ has both the scaled asymptotic negative curvature and the asymptotic covariance for the random slope equal to the Fisher information, that is,
  \[
    \lim_{n\to\infty} -\E\frac{1}{n} \frac{\partial^2 \ell(\theta;Y_{1:n})}{\partial^2 \theta}
    = \lim_{n\to\infty} \frac{1}{n}\text{Var}\left( \frac{\partial \ell(\theta; Y_{1:n})}{\partial \theta} \right) = \mathcal I(\theta_0),
  \]
  the LAN for $\mu(\theta;Y_{1:n})$ has distinct values of $K_1(\theta_0)$ and $K_2(\theta_0)$.
\item Proposition~\ref{prop:lan_sim} considers a local approximation around the simulation-based proxy $\theta_* = \proxy(\theta_0)$.
\end{enumerate}

Ionides et al.\@ \cite{ionides2017monte} overlooked these differences, thereby introducing bias and misspecifying uncertainty.
This paper properly addresses these issues by using a metamodel constructed based on Proposition~\ref{prop:lan_sim}.

\subsection{Simulation metamodel}\label{sec:metamodel}
\begin{algorithm}[t]
  \caption{Simulation-based parameter inference using a metamodel for $\ell^S(\theta;y)$}
  \label{alg:sbi}
  \begin{algorithmic}[1]
    \State \textbf{Input}: observations $y_{1:n} = (y_1, \dots, y_n)$; collection of parameter values $\{\theta_m; m\in 1 \cl M\}$ at which the process $X$ is simulated to obtain log-likelihood estimates
    \State \textbf{Output}: p-values for hypothesis tests on the simulation-based proxy $\theta_*$
    \State For each $m \in 1\cl M$, simulate $X$ under $P_{\theta_m}$ to obtain a simulated log-likelihood $\ell^S(\theta_m;y_{1:n})$
    \State Fit a quadratic polynomial to the points $\{(\theta_m,\, \ell^S(\theta_m, y_{1:n})); m \in 1\cl M\}$ with weights $w_m$\label{line:regression_n}
    \State Estimate $K_1(\theta_0)$ as described in Algorithm~\ref{alg:K1est_block}\;\label{line:K1est}
    \State Carry out a metamodel likelihood ratio test as described in Section~\ref{sec:est_proxy} using the fitted quadratic polynomial and the estimate $\hat K_1$
  \end{algorithmic}
\end{algorithm}

Proposition~\ref{prop:lan_sim} states the local asymptotic property for $\mu(\theta; Y_{1:n}) = \E \ell^S(\theta; Y_{1:n})$ due to the randomness in $Y_{1:n}$ under $P_{\theta_0}^Y$.
We will additionally assume that for given observations $y_{1:n}$, the simulated log-likelihood $\ell^S(\theta; y_{1:n})$ is approximately normally distributed.
The approximate normality of $\ell^S(\theta;y_{1:n})$ can be obtained as a consequence of the central limit theorem when $\ell^S_i(\theta; y_i)$ are independent or weakly dependent.
Mixing conditions similar to those discussed in Section~\ref{sec:supp_mixing} justifying Assumption~\ref{assum:1st_diff} can also justify the asymptotic normality of $\ell^S(\theta;y_{1:n})$.

\begin{assum}\label{assum:mcvar}
  For a given sequence of observations $\{y_i; i\geq 1\}$, the simulated log-likelihoods $\ell^S(\theta;y_{1:n})$ satisfy
  \[
    \frac{\ell^S(\theta; y_{1:n}) - \mu(\theta; y_{1:n})}{\sigma(\theta; y_{1:n})} \Rightarrow \N(0,1)
  \]
  for every $\theta$ in a neighborhood of $\theta_*$ where $\sigma^2(\theta;y_{1:n}) = \Var \{\ell^S(\theta; y_{1:n})\}$.
\end{assum}


Assuming that the LAN for $\mu(\theta;Y_{1:n})$ and Assumption~\ref{assum:mcvar} hold, the simulated log-likelihood $\ell^S(\theta;y_{1:n})$ is approximately normally distributed with a quadratic mean function.
We formulate these facts as our metamodel, consisting of two parts.
The first part states the normality of $\ell^S(\theta; y_{1:n})$ and the quadraticity of $\mu(\theta;y_{1:n})$ for given observations $y_{1:n}$.
This will be referred to as the \emph{conditional simulation metamodel} given the data.

\begin{defn}[Conditional simulation metamodel given data]\label{defn:nlqmeta}
  Given observations $y_{1:n}$, the simulated log-likelihood is distributed as
  \begin{equation}
    \ell^S(\theta;y_{1:n}) \sim \N\left\{a + b^\top \theta + \theta^\top c\theta, \, \frac{\sigma^2(y_{1:n})}{w(\theta)}\right\}
    \label{eqn:nlqmeta}
  \end{equation}
  for some $a \in \R$, $b \in \R^d$, $c \in \R^{d\times d}$, $\sigma^2\in \R$, and a positive function $w(\theta)$, for all $\theta$ in a neighborhood of $\theta_*$.
\end{defn}

Here $w(\theta)$ represents the precision of the simulated log-likelihood $\ell^S(\theta;y_{1:n})$, often signifying the amount of Monte Carlo efforts to obtain $\ell^S(\theta;y_{1:n})$.
For instance, the log of the likelihood estimator obtained through the particle filter has a variance that scales inversely proportional to the number of particles used \cite{berard2014lognormal}.
In this case, $w(\theta)$ is equal to the number of particles.
Section~\ref{sec:sbi_pomp} discusses more on the use of the particle filter for POMP models.
Thus we assume that $w(\theta)$ is known at each $\theta$ where simulations are performed, whereas $\sigma^2(y_{1:n})$ is an unknown quantity that will be estimated via simulations.

The second part of our metamodel describes the statistical properties of the coefficients of the quadratic mean function $\mu(\theta; Y_i)$ explained by the LAN.
Comparing \eqref{eqn:lan} and \eqref{eqn:nlqmeta}, we see that
\begin{equation*}
  \begin{split}
    &\mu\left(\theta_* + \frac{t}{\sqrt n};Y_{1:n}\right) - \mu(\theta_*; Y_{1:n})\\
    &=b^\top \left(\theta_* + \frac{t}{\sqrt{n}}\right) + \left(\theta_* + \frac{t}{\sqrt n}\right)^\top c \left(\theta_* + \frac{t}{\sqrt n}\right) - b^\top \theta_* - \theta_*^\top c \theta_*\\
    &= b^\top \frac{t}{\sqrt n} + \frac{1}{n} t^\top c t + \frac{2}{\sqrt n} \theta_*^\top c t
      = S_n^\top t - \frac{1}{2} t^\top K_2(\theta_0) t + o_p(1)
  \end{split}
\end{equation*}
as $n\to\infty$ under $Y_{1:n} \sim P^Y_{\theta_0}$. 
Thus we have
\begin{equation}
  \begin{split}
    &\frac{1}{n}c = -\frac{1}{2} K_2(\theta_0) + o_p(1),\\
    &\frac{1}{\sqrt n} b = S_n - \frac{2}{\sqrt n} c \theta_* + o_p(1) = S_n + \sqrt n K_2(\theta_0) \theta_* + o_p(1).
  \end{split}
  \label{eqn:meta_K_relationship}
\end{equation}
An asymptotic approximation given by \eqref{eqn:meta_K_relationship} will be referred to as a \emph{marginal simulation metamodel}.

\begin{defn}[Marginal simulation metamodel]\label{defn:marginalmeta}
  The linear and quadratic coefficients $b$ and $c$ for the mean function of the conditional metamodel (Definition~\ref{defn:nlqmeta}) are given by
  \begin{equation}
    c(Y_{1:n}) = -\frac{n}{2} K_2(\theta_0),
    \qquad
    b(Y_{1:n}) \sim \N \left\{nK_2(\theta_0) \theta_*, nK_1(\theta_0) \right\},
    \label{eqn:marginalmeta}
  \end{equation}
  where $Y_{1:n} \sim P^Y_{\theta_0}$.
\end{defn}
We note that we do not make any assumption about the distribution of the constant term $a(Y_{1:n})$.
Since the LAN property concerns the relative difference $\mu(\theta + \tfrac{t}{\sqrt n}; Y_{1:n}) - \mu(\theta; Y_{1:n})$, it enables estimation of $\proxy(\theta_0)$ without any knowledge on the constant term.

Algorithm~\ref{alg:sbi} gives an outline of our simulation-based parameter inference method using the metamodel.
For inference, we simulate the process $X$ under $M$ parameter values $\theta_m$, $m\in 1 \cl M$, and obtain simulated log-likelihoods $\ell^S(\theta_m; y_{1:n})$.
The metamodel parameters $a$, $b$, $c$, and $\sigma^2$ can be estimated by fitting a quadratic polynomial to the points $\{(\theta_m, \ell^S(\theta_m; y_{1:n})); m \in 1\cl M\}$ through standard weighted regression with weights $w(\theta_m)$.
If the regression estimates are given by $(\hat a, \hat b, \hat c)$, we obtain a point estimate for the MESLE by
\[
  \arg\max_\theta \hat a + \hat b^\top \theta + \theta^\top \hat c \theta = -\frac{1}{2} \hat c^{-1} \hat b.
\]
We note that the metamodel parameters $(a,b,c)$ and $\sigma^2$ may not be estimated unless $M \geq \frac{\dimpar(\dimpar+1)}{2} + \dimpar + 1$ where $d$ is the dimension of the parameter vector.


We remark on the appropriate number of simulations $M$.
In order for the MESLE to be identifiable, the information on the location of $\theta_*$ given by the curvature of the quadratic mean function in \eqref{eqn:nlqmeta} should be at least comparable to the uncertainty in the estimated mean function.
This condition can be approximately stated as
\begin{equation}
  (\delta \theta)^\top c(y_{1:n}) (\delta \theta) \gtrsim \frac{\sigma(y_{1:n})}{\sqrt M},
  \label{eqn:sig_noise_ratio}
\end{equation}
where $\delta \theta$ is the half-width of the window of parameter values $\{\theta_1,\dots, \theta_M\}$ in which the simulations are carried out.
Since $c(y_{1:n}) = -\tfrac{n}{2} K_2(\theta_0)$ often scales as $\O(n)$ and $\sigma(y_{1:n}) = \text{Var}^{1/2}\{\ell^S(\theta;y_{1:n})\}$ as $\O(\sqrt n)$, Equation \eqref{eqn:sig_noise_ratio} implies that $\theta_\MESLE$ can be estimated at a scale $\Vert \delta \theta \Vert = \O(n^{-1/2})$ that is comparable to the statistical uncertainty $\Vert \theta_\MESLE(y_{1:n}) - \theta_* \Vert = \O(n^{-1/2})$ provided that $M \gtrsim \O(n)$.
If $M \ll n$, the simulation-based uncertainty swamps the statistical uncertainty.

On the other hand, in order for the quadratic approximation of $\mu(\theta;y_{1:n})$ to be valid on the given interval, the third order term in the Taylor expansion of $\mu(\theta; y_{1:n})$ should be negligible relative to the simulation-based uncertainty of order $\O(\sigma(y_{1:n}) \cdot M^{-1/2}) = \O(n^{1/2} M^{-1/2})$.
Since the third order term scales as $\O(n \cdot \Vert \delta \theta\Vert^3)$, the quadratic approximation is valid at a scale $\Vert \delta \theta \Vert = \O(n^{-1/2})$ if $\O(n \cdot \Vert \delta \theta \Vert^3) = \O(n^{-1/2}) \ll \O(n^{1/2}M^{-1/2})$, or $M \ll \O(n^2)$.
In Section~\ref{sec:autoAdjust}, we introduce an algorithm that automatically reduces the weights of simulations performed at points far from the estimated parameter, ensuring that the cubic term in the approximation is statistically insignificant and maintaining the validity of the LAN approximation.

\begin{algorithm}[t]
  \caption{The bootstrap particle filter for partially observed Markov processes}
  \label{alg:pf}
  \begin{algorithmic}[1]
    \State \textbf{Input}: observations $y_{1:n} = (y_1, \dots, y_n)$; parameter value $\theta$; simulator of the distribution of $X_1$ under $P_\theta$; simulator of the conditional distribution of $X_i|X_{i-1}$ for $i\in 2\cl n$ under $P_\theta$; number of particles $J$
    \State \textbf{Output}: particle approximations to the filtering distributions $X_i|y_{1:i}$ for $i\in 1\cl n$; unbiased likelihood estimator $\hat L(\theta;y_{1:n})$
    \For {$i \in 1\cl n$}
      \If {$i=1$}
        \State Simulate $J$ iid particles $X_i^j$, $j\in 1\cl J$, from the distribution of $X_1$ under $P_\theta$
      \Else 
        \State Simulate $X_i^j$ from $X_i|(X_{i-1} = \tilde X_{i-1}^j)$ for $j\in 1\cl J$ under $P_\theta$
      \EndIf
      \State Compute an estimate of the conditional density $p(y_i | y_{1:i-1}; \theta)$ by
      \[
        \hat L_i(\theta) = \frac{1}{J} \sum_{j=1}^J g_i(y_i | X_i^j; \theta)
      \]
      \State Compute resampling weights
      \[
        w_i^j = g_i(y_i | X_i^j; \theta) / \{J\hat L_i(\theta)\}, ~ j\in 1\cl J
      \]
      \State Resample particles by letting $\tilde X_i^j = X_i^{a_i^j}$ where
      \[
        a_i^j \sim \text{Categorical}(\{1,\dots, J\}, \{w_i^1, \dots, w_i^J\})
      \]
    \EndFor
    \State Compute an unbiased likelihood estimator $\hat L(\theta;y_{1:n}) = \prod_{i=1}^n \hat L_i(\theta)$
  \end{algorithmic}
\end{algorithm}

\subsection{Application to partially observed Markov processes (POMPs)}\label{sec:sbi_pomp}

The particle filter is a class of recursive Monte Carlo algorithms for making inference for hidden Markov models.
The bootstrap particle filter runs by simulating the latent Markov process successively from one observation time to the next, without evaluating the Markov transition density.
An unbiased estimator of the likelihood $L(\theta;y_{1:n})$ can be obtained from the collections of simulated particles.
Thus the bootstrap particle filter can enable inference for implicitly defined, partially observed, Markov processes.
Algorithm~\ref{alg:pf} gives a pseudocode for the bootstrap particle filter, with expressions for the unbiased likelihood estimator $\hat L(\theta)$ and for the estimators $\hat L_i(\theta)$ of the conditional densities $p(y_i|y_{1:i-1};\theta)$, $i\in 1\cl n$.

We use the log of the unbiased likelihood estimator as the simulated log-likelihood $\ell^S(\theta;y_{1:n}) = \log \hat L(\theta;y_{1:n})$ and use the individual simulated log-likelihoods given by $\ell^S_i(\theta; y_i) = \log \hat L_i(\theta)$.
By construction, we have $\ell^S(\theta) = \sum_{i=1}^n \ell^S_i(\theta)$, which is Assumption~\ref{assum:simll_sum}.
B\'erard et al.~\cite{berard2014lognormal} showed that the log of the unbiased likelihood estimator minus the exact log-likelihood converges to a Gaussian limit under certain regularity conditions if the number of particles $J$ increases linearly with the time length $n$:
\begin{equation}
  \ell^S(\theta;y_{1:n}) - \ell(\theta;y_{1:n}) \overset{n\to\infty}{\Longrightarrow} \N\left(-\frac{1}{2} \sigma^2, \sigma^2\right),
  \label{eqn:pf_lognormal}
\end{equation}
where $\sigma^2>0$ is a constant that depends on the model, the observed data, and $\lim_{n\to\infty} \frac{n}{J}$.
This result justifies the asymptotic normality of $\ell^S(\theta; y_{1:n})$ described by the conditional metamodel (Definition~\ref{defn:nlqmeta}) for POMP models.

\subsection{Bias in simulation-based inference}\label{sec:bias}
The Jensen bias $B(\theta; y_{1:n}) = \ell(\theta; y_{1:n}) - \mu(\theta; y_{1:n})$ depends on the tail distribution of the simulated log-likelihood $\ell^S(\theta; y_{1:n})$, as Equation~\eqref{eqn:bias_expansion} suggests.
Thus if $\ell^S(\theta; y_{1:n})$ has a heavy right tail, then the Jensen bias can be large and hard to estimate.
However, a bound on the Jensen bias can be obtained when $\ell^S(\theta; y_{1:n})$ has a light, sub-Gaussian right tail.
Specifically, if 
\[
  \Prob[ \ell^S(\theta) - \mu(\theta) \geq t ] \leq \frac{C'}{\sqrt{2\pi} C} e^{-\frac{t^2}{2C^2}}
\]
for every $t\geq 0$ for some $C, C'>0$, then the Jensen bias is upper bounded by $\frac{C^2}{2} + \log (1+C')$.
Another case where the Jensen bias can be bounded is when each individual simulated log-likelihood $\ell^S_i(\theta)$ is bounded within an interval of width $s_i(\theta)$.
Then the Jensen bias is upper bounded by $\sum_{i=1}^n s_i(\theta)$.
This may be useful if the observation space is compact.
Section~\ref{sec:supp_jensen_bias} in the supplementary text gives proofs for these results.

We develop a bound on the inference bias $\Vert \proxy(\theta_0) - \theta_0 \Vert$ as follows.
Consider the cross entropy $H(\theta_0, \theta) := -\E_{\theta_0} \{ \log p_\theta^Y(Y)\}$ and its second order Taylor approximation around $\theta = \theta_0$, given by
\begin{equation}
  \begin{split}
    q_{-H}(\theta) &:= -H(\theta_0, \theta_0) - \frac{1}{2} (\theta - \theta_0)^\top \mathcal I(\theta_0) (\theta - \theta_0).\\
  \end{split}
  \label{eqn:H_quad_approx}
\end{equation}
Here $\mathcal I(\theta_0)=\left[ \frac{\partial^2}{\partial \theta^2}H(\theta_0, \theta) \right]_{\theta = \theta_0}$ denotes the Fisher information at $\theta_0$. 
Similarly, consider a second order approximation to $U(\theta_0,\theta)$ around $\theta = \theta_*$ given by
\begin{equation}
  \begin{split}
    q_U(\theta) &:= U(\theta_0, \theta_*) - \frac{1}{2} (\theta - \theta_*)^\top \mathcal J(\theta_0, \theta_*) (\theta - \theta_*),
  \end{split}
  \label{eqn:U_quad_approx}
\end{equation}
where $\mathcal J(\theta_0, \theta_*) := -\left[ \frac{\partial^2}{\partial \theta^2}U(\theta_0, \theta) \right]_{\theta = \theta_*}.$
Provided that $\ell^S(\theta)$ is given by the logarithm of an unbiased likelihood estimator, we obtain the following result.
\begin{prop}\label{prop:proxy_bias}
  Suppose that the quadratic approximations $q_{-H}(\theta)$ and $q_U(\theta)$ given by \eqref{eqn:H_quad_approx}--\eqref{eqn:U_quad_approx} are $\epsilon$-accurate on a $\delta$-neighborhood of $\proxy(\theta_0)=\theta_*$, that is,
  \[
    | -H(\theta_0, \theta) - q_{-H}(\theta) | \leq \epsilon, \quad
    | U(\theta_0, \theta) - q_U(\theta) | \leq \epsilon,
  \]
  for every $\theta$ such that $\Vert \theta - \theta_*\Vert \leq \delta$.
  Suppose further that both the Fisher information $\mathcal I(\theta_0)$ and $\J(\theta_0, \theta_*)$ are positive definite and that 
  \[
    \bar B := \sup_{\theta; \Vert \theta - \theta_*\Vert \leq \delta} \E_{Y_{1:n}\sim P_{\theta_0}^Y} B(\theta; Y_{1:n})
  \]
  is finite.
  If the smallest eigenvalue $\lambda$ of $\J(\theta_0,\theta_*)$ satisfies $\lambda \delta^2 \geq 2 (\bar B + 2\epsilon)$, then we have
  \[
    \Vert \theta_0 - \theta_* \Vert \leq 2 \delta^{-1}\lambda^{-1} (\bar B + 2\epsilon).
  \]
\end{prop}
The proof is provided in Section~\ref{sec:supp_proxy_bias} of the supplementary text.
When $n$ is large, $\J(\theta_0,\theta_*)$, and consequently $\lambda$, can be approximately estimated by $-2\hat c$, where $\hat c$ is the quadratic coefficient of the quadratic polynomial fitted to $\{(\theta_m, \ell^S(\theta_m)); m\in 1\cl M\}$.

A similar bound on the difference between the MESLE and the MLE can be established.
\begin{prop}\label{prop:mesle_bias}
  Consider a given observation sequence $y_{1:n}$.
  Suppose that for every $\theta$ such that $\Vert \theta - \theta_\MESLE\Vert \leq \delta$, the following quadratic approximations are $\epsilon$-accurate:
  \begin{gather*}
    \left| \ell(\theta) - \ell(\theta_\MLE) - \frac{1}{2} (\theta - \theta_\MLE)^\top \frac{\partial^2 \ell}{\partial \theta^2}(\theta_\MLE) (\theta - \theta_\MLE) \right| \leq \epsilon, \\
    \left| \mu(\theta) - \mu(\theta_\MESLE) - \frac{1}{2} (\theta - \theta_\MESLE)^\top \frac{\partial^2 \mu}{\partial \theta^2}(\theta_\MESLE) (\theta - \theta_\MESLE) \right| \leq \epsilon.
  \end{gather*}
  Suppose further that both $-\frac{\partial^2 \ell}{\partial \theta^2}(\theta_\MLE)$ and $-\frac{\partial^2 \mu}{\partial \theta^2}(\theta_\MESLE)$ are positive definite and that $\bar B' := \sup_{\theta; \Vert \theta - \theta_\MESLE\Vert \leq \delta} B(\theta; y_{1:n})$ is finite.
  If the smallest eigenvalue $\lambda'$ of $-\frac{\partial^2 \mu}{\partial \theta^2}(\theta_\MESLE)$ satisfies $\lambda' \delta^2 \geq 2 (\bar B' + 2\epsilon)$, then we have
  \[
    \Vert \theta_\MLE - \theta_\MESLE \Vert \leq 2 \delta^{-1} {\lambda'}^{-1} (\bar B' + 2\epsilon).
  \]
\end{prop}

\section{Estimation of $\theta_\MESLE$ and $K_1(\theta_0)$}\label{sec:est_mesle_K1}

\subsection{Estimation of the metamodel parameters}\label{sec:est_metaparam}
For our normal, locally quadratic metamodel
\[
  \ell^S(\theta;y_{1:n}) \sim \N\left\{ a + b^\top\theta + \theta^\top c \theta, \frac{\sigma^2(y_{1:n})}{w(\theta)} \right\},
\]
the mean function $a+ b^\top \theta + \theta^\top c \theta$ can be estimated by weighted least square regression.
We will write $A = (a, b^\top, \vech(c)^\top)^\top$ where $\vech(c)$ is the half-vectorization of the symmetric matrix $c=(c_{ij})_{i\in 1:d, j\in 1:d}$, given by 
\[
  \vech(c) = (c_{11}, c_{21}, \dots, c_{\dimpar 1}, c_{22}, \dots, c_{\dimpar 2}, \dots, c_{\dimpar \dimpar})^\top \in \R^{\frac{d(d+1)}{2}}.
\]
We will write $\theta = (\theta_{(1)}, \dots, \theta_{(\dimpar)})^\top$ and denote by $\theta^2$ the $\dimpar\times\dimpar$ matrix whose $k$-th diagonal entry is $\theta_{(k)}^2$ and whose $(k,l)$-th entry is $2\theta_{(k)}\theta_{(l)}$ for $k\neq l$.
We can then write
\[
  \theta^\top c \theta = \vech(\theta^2)^\top \vech(c)
\]
and the mean function as
\[
  \mu(\theta; A) := a + b^\top \theta + \theta^\top c \theta
  = (1, \theta^\top, \vech(\theta^2)^\top) A
  =: {\theta^{0:2}}^\top A
\]
where $\theta^{0:2} = (1, \theta^\top, \vech(\theta^2)^\top)^\top \in \R^{\frac{d^2+3d+2}{2}}$.
We will write
\[
  \theta^{0:2}_{1:M} = \begin{pmatrix} {\theta_1^{0:2}}^\top \\ \vdots \\ {\theta_M^{0:2}}^\top \end{pmatrix} \in \R^{M\times \left(\frac{d^2+3d+2}{2}\right)}.
\]

The log-likelihood for $A$ and $\sigma^2$ under our conditional metamodel is given by
\[
  \log p_\text{meta}(\ell^S(\theta_{1:M}) | A, \sigma^2)
  = \sum_m -\frac{1}{2} \log \left(2\pi\frac{\sigma^2}{w_m}\right) - \frac{(\ell^S(\theta_m) - \mu(\theta_m;A))^2}{2\sigma^2/w_m}
\]
where $w_m := w(\theta_m)$.
Maximizing this metamodel log-likelihood is equivalent to a quadratic regression with weights $w_m$, $m\in 1\cl M$.
We will write $\ell^S_m := \ell^S(\theta_m)$ and denote by $W$ be the diagonal matrix with diagonal entries $w_1, \dots, w_M$.
We assume that $\theta^{0:2}_{1:M} \in \R^{M\times \left(\frac{d^2+3d+2}{2}\right)}$ has rank $\frac{d^2+3d+2}{2}$.
We use the notation $\Vert v \Vert^2_W := v^\top W v$ such that
\[
  \sum_{m=1}^M w_m \cdot (\ell^S(\theta_m) - \mu(\theta_m;A))^2 = \Vert \ell^S_{1:M} - {\theta^{0:2}_{1:M}}^\top A \Vert_W^2
\]
where $\ell^S_{1:M} := (\ell^S(\theta_1), \dots, \ell^S(\theta_M))^\top$.
The maximum metamodel likelihood estimates for $A$ and $\sigma^2 = \sigma^2(y_{1:n})$ are given by
\begin{gather}
  \hat A = (\hat a, \hat b^\top, \vech(\hat c)^\top)^\top = \{{\theta^{0:2}_{1:M}}^\top W \theta^{0:2}_{1:M}\}^{-1} {\theta^{0:2}_{1:M}}^\top W \ell^S_{1:M}, \label{eqn:I_Ahat}\\
  \hat \sigma^2 = \frac 1 M \Vert \ell^S_{1:M} - {\theta^{0:2}_{1:M}} \hat A\Vert_W^2.  \label{eqn:I_sigmahat}
\end{gather}
For the test of hypotheses
\[
  H_0: A = A_0 = (a_0,b_0^\top,\vech(c_0)^\top)^\top, \ \sigma^2 = \sigma_0^2, \qquad H_1: \text{ not }H_0,
\]
the metamodel log-likelihood ratio statistic is given by
\begin{equation}
  \ml{MLLR}_{A_0,\sigma_0^2}
  := \frac{\log p_\text{meta}(\ell^S_{1:M} | A_0, \sigma^2_0)}{\sup_{A,\sigma^2} \log p_\text{meta}(\ell^S_{1:M} | A, \sigma^2)}
  = \frac{M}{2} \log \frac{\hat\sigma^2}{\sigma_0^2} - \frac{\Vert \ell^S_{1:M} - {\theta^{0:2}_{1:M}} A_0 \Vert_{W}^2}{2\sigma_0^2} + \frac{M}{2}.
  \label{eqn:MLLR_M1}
\end{equation}

As the number of simulation $M$ grows, $-2 \cdot \ml{MLLR}_{A_0, \sigma_0^2}$ converges in distribution to $\chi_{\frac{d^2+3d+4}{2}}^2$ under the null hypothesis $H_0: A=A_0, \sigma^2 = \sigma^2_0$.
The degrees of freedom $\frac{d^2+3d+4}{2}$ equals the number of scalar metamodel parameters, namely $a$, $d$ entries of $b$, $\tfrac{d^2+d}{2}$ diagonal and lower triangular entries of $c$, and $\sigma^2$.
The finite sample distribution of the $\ml{MLLR}_{A_0, \sigma_0^2}$ statistic under the null hypothesis is characterized in supplementary Section~\ref{sec:proofs_est_metaparam}.

\subsection{Estimation and uncertainty quantification for the MESLE}\label{sec:est_mesle}
The maximum metamodel likelihood estimator for $\theta_\MESLE$ is given by
\[
  \hat \theta_\MESLE = -\frac{1}{2} \hat c^{-1} \hat b,
\]
where $\hat b$ and $\hat c$ are obtained by quadratic regression, \eqref{eqn:I_Ahat}.
The variance of $\hat \theta_\MESLE$ scales as $\O(M^{-1}n^{-1})$.
This can be seen by using the delta method as follows \cite{vandervaart1998asymptotic}.
From \eqref{eqn:I_Ahat}, we can show that 
\begin{equation}
  \text{Var}(\hat A) = \sigma^2 \{ {\theta_{1:M}^{0:2}}^\top W \theta_{1:M}^{0:2} \}^{-1}
  \label{eqn:varAhat}
\end{equation}
and
\[
  \sqrt M(\hat \theta_\MESLE - \theta_\MESLE) \Rightarrow \N(0, \Sigma_\MESLE),
\]
where
\[
  \Sigma_\MESLE = \sigma^2 \cdot \left( \frac{\partial (-\frac{1}{2}c^{-1}b)}{\partial (a,b,c)} \right) \left( \lim_{M\to\infty} M \{ {\theta_{1:M}^{0:2}}^\top W \theta_{1:M}^{0:2} \}^{-1} \right) \left( \frac{\partial (-\frac{1}{2}c^{-1}b)}{\partial (a,b,c)} \right)^\top.
\]
According to our marginal metamodel, $b$ and $c$ scale as $\O(n)$, and we often have $\sigma^2 = \O(n)$.
We also have ${\theta^{0:2}_{1:M}}^\top W \theta^{0:2}_{1:M} = \O(M)$.
Hence, we have $\text{Var}(\hat\theta_\MESLE) = M^{-1} \cdot \O(n^{-2} \sigma^2) = \O(M^{-1}n^{-1})$.

Now we consider a finite sample test on $\theta_\MESLE$ without using the delta method.
The metamodel log-likelihood ratio statistic for 
\begin{equation}
  H_0: \theta_\MESLE(y_{1:n}) = \theta_{H_0}, \qquad H_1: \theta_\MESLE(y_{1:n}) \neq \theta_{H_0},
  \label{eqn:ht_mesle}
\end{equation}
is given by
\begin{equation}
  \ml{MLLR}_{\theta_{H_0}} = \sup\left\{\ml{MLLR}_{A_0,\sigma_0^2} ; \, -\frac{1}{2}c_0^{-1} b_0 = \theta_{H_0}, \, \sigma_0^2>0\right\}.
  \label{eqn:mllr_MESLE}
\end{equation}
For a given parameter vector $\theta\in \R^\dimpar$, we denote by $\theta_\matr$ a $\dimpar \times \frac{\dimpar(\dimpar+1)}{2}$ matrix
\begin{equation}
  \theta_\matr
  := \begin{pmatrix}
      \theta_{(1)} & \theta_{(2)} & \cdots       & \cdots & \theta_{(\dimpar)} &              &              &        &                    &              &        &        &                    \\
                   & \theta_{(1)} &              &        &                    & \theta_{(2)} & \theta_{(3)} & \cdots & \theta_{(\dimpar)} &              &        &        &                    \\
                   &              & \theta_{(1)} &        &                    &              & \theta_{(2)} &        &                    & \theta_{(3)} & \cdots &        &                    \\
                   &              &              & \ddots &                    &              &              & \ddots &                    &              & \ddots & \cdots &                    \\
                   &              &              &        & \theta_{(1)}       &              &              &        & \theta_{(2)}       &              &        & \cdots & \theta_{(\dimpar)} 
  \end{pmatrix}
  \label{eqn:mat_notation}
\end{equation}
so that we can write $c\theta = \theta_\matr \vech(c)$.
The test statistic and the critical value are given as follows.
\begin{prop}\label{prop:mllr_MESLE}
  Let
  \[
    U = \begin{pmatrix} u_{aa} & \mathbf u_{a, bc} \\ \mathbf u_{bc,a} & U_{bc,bc} \end{pmatrix}
    := {\theta^{0:2}_{1:M}}^\top W \theta^{0:2}_{1:M}
  \]
  where $u_{aa} \in \R$ and $U_{bc,bc} \in \R^{\frac{d^2+3d}{2} \times \frac{d^2+3d}{2}}$,
  and let
  \[
    V := U_{bc,bc} - \mathbf u_{bc, a} u_{aa}^{-1} \mathbf u_{a,bc}.
  \]
  Then the $\ml{MLLR}_{\theta_{H_0}}$ statistic \eqref{eqn:mllr_MESLE} is given by
  \[
    \ml{MLLR}_{\theta_{H_0}} = -\frac{M}{2}\log\left(\frac{\xi}{M\hat\sigma^2} + 1\right)
  \]
  where
  \[
    \xi := (\hat b + 2 \hat c \theta_{H_0})^\top 
    \left\{ \begin{pmatrix} I_\dimpar \\ 2\theta_{H_0,\matr}^\top \end{pmatrix}^\top V^{-1} \begin{pmatrix} I_\dimpar \\ 2\theta_{H_0,\matr}^\top \end{pmatrix} \right\}^{-1}
    (\hat b + 2 \hat c \theta_{H_0}).
  \]
  Under the null hypothesis $H_0: \theta_\MESLE = \theta_{H_0}$, namely if $A=A_0 = (a_0, b_0^\top, \vech(c_0)^\top)^\top$ satisfies $-\frac{1}{2} c_0^{-1} b_0 = \theta_{H_0}$, we have for any $\sigma^2 > 0$
  \[
    \frac{(M-\frac{d^2+3d+2}{2})\xi}{M d \hat\sigma^2} \sim F_{d,M-\frac{d^2+3d+2}{2}}.
  \]
\end{prop}
Proposition~\ref{prop:mllr_MESLE} suggests rejecting the null hypothesis in \eqref{eqn:ht_mesle} when 
\[
  \frac{(M-\frac{d^2+3d+2}{2})\xi}{M d \hat\sigma^2} > F_{d,M-\frac{d^2+3d+2}{2},\alpha}
\]
to achieve a significance level of $\alpha$.
The p-value is given by
\[
  \Prob[F_{d,M-\frac{d^2+3d+2}{2}}> (M-\frac{d^2+3d+2}{2})\xi/(Md\hat\sigma^2)].
\]
For the case where $\theta$ is one dimensional (i.e., $d=1$), a confidence interval for $\theta_\MESLE$ can be obtained as follows.
\begin{corollary}\label{cor:ci_mesle}
  Assume that the parameter space is one dimensional (i.e., $\mathsf\Theta \subseteq \R$).
  Let $V$ be as in Proposition~\ref{prop:mllr_MESLE}.
  Then a level $1-\alpha$ confidence interval for $\theta_\MESLE$ is given by
  \begin{multline}
    \big\{ \theta ;
      \big[4(M-3)\hat c^2 \det V - 4M\hat\sigma^2 F_{1,M-3,\alpha} V_{bb} \big] \theta^2 \\
      +\big[4(M-3)\hat b \hat c \det V + 4M\hat \sigma^2 F_{1,M-3,\alpha}V_{bc} \big] \theta \\
      + (M-3)\hat b^2\det V - M\hat \sigma^2 F_{1,M-3,\alpha} V_{cc} < 0 \big\}.
      \label{eqn:ci_MESLE}
  \end{multline}
\end{corollary}
If the quadratic coefficient in \eqref{eqn:ci_MESLE} is negative and the discriminant of the left hand side is nonnegative, then the constructed confidence interval will be of the form $(-\infty, \text{LB}) \cup (\text{UB}, \infty)$ where $-\infty<\text{LB}<\text{UB}<\infty$.
This situation arises roughly when $\hat c^2 \lesssim \hat\sigma^2 / M$.
To see this, we note that for $d=1$ we have
\[
  V = \begin{pmatrix} V_{bb} & V_{bc} \\ V_{bc} & V_{cc} \end{pmatrix}
  = (\sum_m w_m) \begin{pmatrix} \overline{\theta^2} - (\overline\theta)^2 & \overline{\theta^3} - \overline\theta\cdot \overline{\theta^2} \\ \overline{\theta^3} - \overline{\theta}\cdot \overline{\theta^2} & \overline{\theta^4} - (\overline{\theta^2})^2 \end{pmatrix}
\]
where $\overline{\theta^j} = (\sum_m w_m \theta_m^j)/(\sum_m w_m)$, $j\in 1\cl 4$, and thus $\det V = V_{bb}V_{cc} - V_{bc}^2 = \O(M) \cdot V_{bb}$.
The coefficient for $\theta^2$ in \eqref{eqn:ci_MESLE} may be negative if $\hat c^2 \lesssim \hat \sigma^2 / M$.
This implies that the MESLE may not be estimated if the signal to noise ratio is too small.
If the discriminant of the quadratic polynomial in \eqref{eqn:ci_MESLE} is negative, the confidence interval is the entire real line $(-\infty, \infty)$, and the MESLE may not be estimated either.

The p-value and the confidence interval developed in this section can be numerically found using the \texttt{ht} and \texttt{ci} functions in \textsf{R} package \texttt{sbim} (\url{https://CRAN.R-project.org/package=sbim}).

\subsection{Estimation of $K_1(\theta_0)$}\label{sec:K1est}
The marginal metamodel (Defnition~\ref{defn:marginalmeta}) implies that $K_2(\theta_0)$ can be estimated by $-\frac{2}{n} \hat c$, where $\hat c$ is given by \eqref{eqn:I_Ahat}.
Estimation of $K_1(\theta_0)$ is more complicated, because it involves estimating the variability in $\tfrac{\partial \mu}{\partial \theta}(\theta_*; Y_{1:n})$, where only a single set of observations $y_{1:n}$ is available.
One possible method for estimating $K_1(\theta_0)$ is to use parametric bootstrap, where we generate $n_\text{boot}$ sets of observations $y_{1:n}^{\text{boot},k}$ at $\theta = \hat\theta_\MESLE$ for $k\in 1\cl n_\text{boot}$ and compute the sample covariance matrix of $\tfrac{\partial \hat\mu}{\partial\theta}(\hat\theta_\MESLE; y_{1:n}^{\text{boot},k})$ for those bootstrap samples.
However, this approach has two major disadvantages.
First, for each bootstrap sample, simulated log-likelihoods $\ell^S(\theta_m; y_{1:n}^{\text{boot},k})$ need to be obtained in order to estimate $\frac{\partial\hat\mu}{\partial\theta}(\hat\theta_\MESLE; y_{1:n}^{\text{boot},k})$, which increases the computational complexity by a factor of $n_\text{boot}$.
Another disadvantage is that if our simulator $P_\theta$ is misspecified, the sample variance of $\tfrac{\partial \hat\mu}{\partial \theta}(\hat\theta_\MESLE; y_{1:n}^{\text{boot},k})$ may have a non-negligible bias. 

Here we propose a method where $K_1(\theta_0)$ is estimated using the given data $y_{1:n}$.
We divide the observations $y_{1:n}$ into blocks and assume that these blocks of observations are approximately independent of each other and identically distributed, marginally over the law of $X$.
This assumption may hold if the process $X$ is approximately stationary and possesses a mixing property.
Since the derivative of the mean function evaluated at $\theta_*$ is given by
\[
  \frac{\partial \mu}{\partial \theta}(\theta_*;A)
  = b + 2c\theta_*
  = (\mathbf 0_\dimpar, I_\dimpar, 2\theta_{*,\matr}) A,
\]
where $\mathbf 0_\dimpar \in \R^\dimpar$ is a column vector of zeros, the derivative of the mean function at $\theta_*$ can be estimated by
\[
  \widehat{\frac{\partial \mu}{\partial \theta}}(\theta_*;Y_{1:n})
  = \frac{\partial \hat \mu}{\partial \theta}(\theta_*; \hat A) 
  = \hat b + 2 \hat c\theta
  = (\mathbf 0_\dimpar, I_\dimpar, 2\theta_{*,\matr}) \hat A.
\]
where $\hat A = (\hat a, \hat b, \hat c)$ is given by Equation~\eqref{eqn:I_Ahat}.
Since $\hat A$ is unbiased for $A$, we have
\[
  \E \left( \widehat{\frac{\partial \mu}{\partial\theta}}(\theta_*;Y_{1:n}) \middle| Y_{1:n}\right)
  = b + 2c\theta_*
  = \frac{\partial \mu}{\partial \theta}(\theta_*;Y_{1:n}).
\]

\begin{algorithm}[t]
  \caption{Estimation of $K_1(\theta_0)$ using block partitioning}
  \label{alg:K1est_block}
  \begin{algorithmic}[1]
    \State Choose $\vartheta$, an approximate guess for $\theta_*$, such as the average of $\{\theta_m; m \in 1\cl M\}$
    \State Fit a quadratic polynomial to $\{(\theta_m, \ell^S(\theta_m; y_{1:n})); m\in 1\cl M\}$ and let $\widehat{\frac{\partial \mu}{\partial \theta}}(\vartheta; y_{1:n})$ be the slope of the fitted quadratic polynomial at $\vartheta$ \label{line:K1_regression_n} 
    \State Obtain an estimate of $\frac{1}{n}\E\left\{\Var\left(\widehat{\frac{\partial\mu}{\partial\theta}}(\vartheta;Y_{1:n})\middle| Y_{1:n}\right)\right\}$ via \eqref{eqn:K1_2nd_term}\label{line:exp_condvar}
    \State Partition $\{1,\dots, n\}$ into blocks $B_1,\dots, B_K$ such that $(Y_i; i\in B_k)$ are approximately independent of each other 
    \For {$k \in 1\cl K$}
      \State Fit a quadratic polynomial to $\{(\theta_m,\, \sum_{i\in B_k} \ell^S_i(\theta_m; y_i)); m \in 1\cl M\}$\;
      \State Let $\widehat {\frac{\partial \mu_{B_k}}{\partial \theta}}(\vartheta;y_{B_k})$ denote the slope of the fitted quadratic polynomial at $\vartheta$\;\label{line:slope_i_block}
    \EndFor
    \State Let $\frac{1}{n}\widehat{\Var}_{Y_{1:n}\sim P^Y_{\theta_0}}\left\{\widehat{\frac{\partial \mu}{\partial \theta}}(\vartheta; Y_{1:n})\right\}$ be the (weighted) sample variance of $\{\widehat {\frac{\partial \mu_{B_k}}{\partial \theta}}(\vartheta; y_{B_k}); k\in 1\cl K\}$ as given by \eqref{eqn:block_sample_variance}\;\label{line:var_est_block}
    \State Let $\hat K_1(\theta_0)$ be \eqref{eqn:block_sample_variance} minus \eqref{eqn:K1_2nd_term}
  \end{algorithmic}
\end{algorithm}

We can estimate $K_1(\theta_0) = \lim_{n\to\infty} \frac{1}{n} \Var_{Y_{1:n}\sim P^Y_{\theta_0}} \frac{\partial \mu}{\partial \theta}(\theta_*; Y_{1:n})$ using the expression
\begin{multline}
  \frac{1}{n} \text{Var}_{Y_{1:n}\sim P^Y_{\theta_0}} \left\{ \E\left(\widehat{\frac{\partial \mu}{\partial \theta}}(\theta_*; Y_{1:n}) \middle| Y_{1:n} \right)\right\}\\
  = \frac{1}{n} \Var_{Y_{1:n}\sim P^Y_{\theta_0}}\left\{ \widehat{\frac{\partial \mu}{\partial \theta}}(\theta_*; Y_{1:n}) \right\}
  - \frac{1}{n} \E_{Y_{1:n}\sim P^Y_{\theta_0}}\left\{ \Var\left( \widehat{\frac{\partial \mu}{\partial \theta}}(\theta_*; Y_{1:n}) \middle| Y_{1:n} \right) \right\},
  \label{eqn:K1_approx}
\end{multline}
The steps for estimating $K_1(\theta_0)$ are summarized in Algorithm~\ref{alg:K1est_block}.
The conditional variance in the second term on the right hand side of \eqref{eqn:K1_approx} is given by
\[
  \Var\left( \widehat{\frac{\partial \mu}{\partial \theta}}(\theta_*; Y_{1:n}) \middle | Y_{1:n} \right)
  = (\mathbf 0_\dimpar, I_\dimpar, 2\theta_{*,\matr}) \Var(\hat A | Y_{1:n}) (\mathbf 0_\dimpar, I_\dimpar, 2\theta_{*,\matr})^\top,
\]
where we have
\[
  \Var(\hat A|Y_{1:n}) = ({\theta^{0:2}_{1:M}}^\top W \theta^{0:2}_{1:M})^{-1} \sigma^2(Y_{1:n}).
\]
Thus, the second term on the right hand side of \eqref{eqn:K1_approx} can be approximated by
\begin{equation}
  \frac{\hat\sigma^2(Y_{1:n})}{n} (\mathbf 0_\dimpar, I_\dimpar, 2\vartheta_{\matr}) \{{\theta^{0:2}_{1:M}}^\top W \theta^{0:2}_{1:M}\}^{-1} (\mathbf 0_\dimpar, I_\dimpar, 2\vartheta_{\matr})^\top,
  \label{eqn:K1_2nd_term}
\end{equation}
where $\vartheta$ is a reasonable guess for $\theta_*$, such as the average of $\theta_1,\dots, \theta_M$, and $\hat \sigma^2(Y_{1:n})$ is given by \eqref{eqn:I_sigmahat}.

The first term on the right hand side of \eqref{eqn:K1_approx}, $\frac{1}{n} \Var_{Y_{1:n}\sim P^Y_{\theta_0}}\left\{ \widehat{\frac{\partial \mu}{\partial \theta}}(\theta_*; Y_{1:n})\right\}$ can be estimated as follows.
Let $B_1, \dots, B_K$ be contiguous, non-overlapping blocks partitioning the set $\{1,\dots,n\}$ chosen such that $\{Y_i; i\in B_k\}$, $k\in 1\cl K$, are approximately independent of each other.
A quadratic polynomial is fitted to $\{(\theta_m, \sum_{i \in B_k} \ell^S_i(\theta_m; y_i); m\in 1\cl M\}$ for each block $k\in 1 \cl K$.
Let the estimated slope of the fitted quadratic polynomial for the $k$-th block evaluated at $\vartheta$ be denoted by $\widehat{\frac{\partial \mu_{B_k}}{\partial \theta}}(\vartheta)$.
Then the first term on the right hand side of \eqref{eqn:K1_approx} can be estimated by the weighted sample variance of $\{\widehat{\frac{\partial \mu_{B_k}}{\partial \theta}}(\vartheta); k\in 1\cl K\}$, given by
\begin{equation}
  \frac{1}{K-1} \sum_{k=1}^K |B_k| \left(\frac{1}{|B_k|}\widehat{\frac{\partial \mu_{B_k}}{\partial \theta}}(\vartheta) - \frac{1}{n} \sum_{k=1}^K \widehat{\frac{\partial \mu_{B_k}}{\partial \theta}}(\vartheta)\right) \left(\frac{1}{|B_k|}\widehat{\frac{\partial \mu_{B_k}}{\partial \theta}}(\vartheta) - \frac{1}{n} \sum_{k=1}^K \widehat{\frac{\partial \mu_{B_k}}{\partial \theta}}(\vartheta)\right)^\top
  \label{eqn:block_sample_variance}
\end{equation}
where $|B_k|$ denotes the size of $B_k$.
An estimate of $K_1(\theta_0)$ is obtained by subtracting \eqref{eqn:K1_2nd_term} from \eqref{eqn:block_sample_variance}.
The standard error of this estimate scales as $\O(n^{-1/2})$.

\section{Estimation and uncertainty quantification for the simulation-based proxy $\theta_*$}\label{sec:est_proxy}
In this section, we develop a hypothesis testing procedure for the simulation-based proxy $\proxy(\theta_0) =: \theta_*$.
We assume that the variance $\sigma^2(\theta; y_{1:n})$ of $\ell^S(\theta;y_{1:n})$ is constant in the local neighborhood of $\theta_*$ where simulations are carried out.
Our approach differs from the method proposed by Ionides et al.\@ \cite{ionides2017monte} in two key ways: (a) we account for the difference between the local asymptotic normality for the log-likelihood function $\ell(\theta;Y)$ and that of the expected simulated log-likelihood function $\mu(\theta;Y)$, and (b) we avoid using the delta method approximation.

We employ a restricted maximum likelihood (REML) approach, using the relative differences $\ell(\theta_m;Y_{1:n}) - \ell(\theta_1;Y_{1:n})$, for $m\in 2\cl M$, to eliminate the dependence on $a(Y_{1:n})$, which often has an intractable distribution. 
The reference point is chosen to $\theta_1$ without loss of generality.
The conditional metamodel (Definition~\ref{defn:nlqmeta}) gives that
\[
    \mu(\theta_m; Y_{1:n}) - \mu(\theta_1; Y_{1:n})
    = b^\top (\theta_m - \theta_1) + \theta_m^\top c \theta_m - \theta_1^\top c \theta_1
    = (\theta_m^{1:2} - \theta_1^{1:2})^\top \begin{pmatrix} b \\ \vech(c) \end{pmatrix},
\]
where $\theta^{1:2}_m = (\theta_m^\top, \vech(\theta^2_m)^\top)^\top \in \R^{M\times \frac{d^2+3d}{2}}$.
We will denote the relative differences by $(\ell^S_2 - \ell^S_1, \cdots, \ell^S_M - \ell^S_1)^\top = C \ell^S_{1:M}$, where $C := (-\mathbf 1_{M-1}, I_{M-1}) \in \R^{(M-1)\times M}$.
The distribution of $C\ell^S_{1:M}$ conditional on the observations $Y_{1:n}$ is given by
\begin{equation}
  C \ell^S_{1:M} | Y_{1:n}
  \sim \N\left( C \theta^{1:2}_{1:M} \begin{pmatrix} b \\ \vech(c) \end{pmatrix}, \, \sigma^2 C W^{-1} C^\top \right).
  \label{eqn:C_ell_LAN}
\end{equation}
We will assume that $C \theta^{1:2}_{1:M}$ has a full column rank. 
The marginal metamodel (Definition~\ref{defn:marginalmeta}) states that
\[
  c(Y_{1:n}) = -\frac{n}{2} K_2(\theta_0),
  \qquad
  b(Y_{1:n}) \sim \N \left\{nK_2(\theta_0) \theta_*, nK_1(\theta_0) \right\} = \N\left\{ -2c\theta_*, nK_1(\theta_0)\right\}.
\]
We will assume that $b(Y_{1:n})$ and $\sigma^2(Y_{1:n})$ are independent.
Under this assumption, we have
\begin{equation}
  C\ell^S_{1:M} | \sigma^2
  \sim \N\left(C \theta^{1:2}_{1:M} \begin{pmatrix} -2c \theta_* \\ \vech(c) \end{pmatrix},\, \sigma^2 C W^{-1} C^\top + C\theta_{1:M} n K_1 \theta_{1:M}^\top C^\top \right).
  \label{eqn:Cl_dist}
\end{equation}
The density of \eqref{eqn:Cl_dist} evaluated for $C\ell^S_{1:M}$ will be referred to as the marginal metamodel likelihood and denoted by $p_\text{meta}$.
Mathematical details for this section are given in the supplementary text Section~\ref{sec:proofs_lan}.
By maximizing this marginal metamodel likelihood using the plug-in estimates $\hat \sigma^2$ given by \eqref{eqn:I_sigmahat} and $\hat K_1$ obtained via Algorithm~\ref{alg:K1est_block}, we obtain point estimators $\hat \theta_*$ and $\hat c$ satisfying
\begin{equation}
  \begin{pmatrix} -2 \hat c \hat\theta_* \\ \vech(\hat c) \end{pmatrix} 
  = \left\{ {\theta_{1:M}^{1:2}}^\top \hat \cqc \theta_{1:M}^{1:2} \right\}^{-1} {\theta_{1:M}^{1:2}}^\top \hat\cqc \ell^S_{1:M},
  \label{eqn:K_theta_second_stage}
\end{equation}
where 
\[
  \hat \cqc := C^\top \{C W^{-1} C^\top + \hat \sigma^{-2} C\theta_{1:M} n \hat K_1 \theta_{1:M}^\top C^\top\}^{-1} C.
\]
The estimate $\hat \theta_*$ was very close to $\hat \theta_\MESLE$ for all numerical examples we considered in Section~\ref{sec:numerical}.
We also obtain a second-stage estimate estimate for the variance $\sigma^2$, given by
\begin{equation}
  \hat\sigma^2_\text{2nd} := \frac{1}{M-1} \left\Vert \ell^S_{1:M} - \theta_{1:M}^{1:2} \left\{ {\theta_{1:M}^{1:2}}^\top \hat \cqc \theta_{1:M}^{1:2} \right\}^{-1} {\theta_{1:M}^{1:2}}^\top \hat\cqc \ell^S_{1:M} \right\Vert_{\hat\cqc}^2.
  \label{eqn:errvar_LAN}
\end{equation}

We carry out a test on the simulation-based proxy
\[
  H_0: \theta_*= \theta_{*,0}, \ H_1: \theta_* \neq \theta_{*,0}
\]
using the marginal metamodel log likelihood ratio defined as
\[
  \ml{MLLR}_{\theta_{*,0}} = \log\frac{\sup_{c,\sigma^2} p_\text{meta}(C\ell^S_{1:M} | \theta_{*,0}, c, \sigma^2)}{\sup_{\theta_*, c, \sigma^2} p_\text{meta}(C\ell^S_{1:M} | \theta_*, c, \sigma^2) }.
\]
The distribution of $\ml{MLLR}_{\theta_{*,0}}$ under $H_0$ is described by Proposition~\ref{prop:MLLR_LAN}.
\begin{prop}\label{prop:MLLR_LAN}
  Let $T(\theta_{*,0}) = \theta_{1:M}^{1:2} \begin{pmatrix}\theta_{*,0,\matr}\\-\frac{1}{2} I_{\frac{d^2+d}{2}} \end{pmatrix}$ and 
  \begin{equation*}
    \proj(\theta_{*,0}) = T(\theta_{*,0}) \{ T(\theta_{*,0})^\top \hat\cqc T(\theta_{*,0}) \}^{-1} T(\theta_{*,0})^\top.
  \end{equation*}
  Then the metamodel log-likelihood ratio $\ml{MLLR}_{\theta_{*,0}}$ is given by
  \[
    \ml{MLLR}_{\theta_{*,0}} = -\frac{M-1}{2} \log \frac{\Vert \{I_{M-1}-\proj(\theta_{*,0})\hat\cqc\} \ell^S_{1:M} \Vert_{\hat\cqc}^2}{\hat\sigma^2_\text{2nd}}.
  \]
  Furthermore, if Assumptions~\ref{assum:simll_sum}-\ref{assum:mcvar} hold, 
  \begin{equation}
    \frac{M-\frac{d^2+3d+2}{2}}{d} \left\{ \frac{\Vert \{I_{M-1}-\proj(\theta_{*,0})\hat\cqc\} \ell^S_{1:M} \Vert_{\hat\cqc}^2}{(M-1)\hat\sigma^2_\text{2nd}}-1 \right\} \sim F_{d,M-\frac{d^2+3d+2}{2}}
    \label{eqn:lan_Fdist}
  \end{equation}
  under $H_0: \theta_* = \theta_{*,0}$,  provided that $\hat\sigma^2 = \sigma^2$ and $\hat K_1 = K_1$.
\end{prop}

Proposition~\ref{prop:MLLR_LAN} suggests that we reject $H_0: \theta_* = \theta_{*,0}$ when
\begin{equation}
  \left\Vert \{I_{M-1}-\proj(\theta_{*,0})\hat\cqc\} \ell^S_{1:M} \right\Vert_{\hat\cqc}^2
  > (M-1)\hat\sigma^2_\text{2nd} \left(\frac{d \cdot F_{d,M-\frac{d^2+3d+2}{2},\alpha}}{M-\frac{d^2+3d+2}{2}}+1\right)
  \label{eqn:proxy_nonreject}
\end{equation}
to achieve an approximate significance level $\alpha$.
If we denote the left hand side of \eqref{eqn:lan_Fdist} by $F$, then an approximate p-value is given by $\Prob[F_{d,M-\frac{d^2+3d+2}{2}}>F]$.

When $d=1$, an approximate level $1-\alpha$ confidence interval for $\theta_*$ can be obtained by inverting the hypothesis test.
\begin{corollary}\label{cor:lan}
  Assume that the parameter space is one dimensional (i.e., $\mathsf\Theta \subseteq \R$).
  Write ${\theta_{1:M}^{1:2}}^\top \hat\cqc \theta_{1:M}^{1:2} = \begin{pmatrix} \rho_{11} & \rho_{12}\\ \rho_{12} & \rho_{22} \end{pmatrix}$ and 
  \[
    \zeta_0 = \left\Vert \ell^S_{1:M} \right\Vert_{\hat\cqc}^2 - (M-1)\hat\sigma^2_\text{2nd} \left(\frac{F_{1,M-3,\alpha}}{M-3} + 1 \right), \qquad
    \begin{pmatrix}\zeta_1 \\ \zeta_2\end{pmatrix}
    = {\theta_{1:M}^{1:2}}^\top \hat\cqc \ell^S_{1:M}.
  \]
  Then under Assumptions~\ref{assum:simll_sum}-\ref{assum:mcvar}, an approximate level $1-\alpha$ confidence interval for $\theta_*$ is given by
  \begin{equation}
    \left\{\theta_{*,0};\, (\zeta_0 \rho_{11} - \zeta_1^2)\theta_{*,0}^2 + (\zeta_1 \zeta_2 - \zeta_0 \rho_{12})\theta_{*,0} + \frac{1}{4}(\rho_{22}\zeta_0 - \zeta_2^2) < 0 \right\}.
    \label{eqn:ci_sbs}
  \end{equation}
\end{corollary}
The approximate p-value and the approximate confidence interval developed in this section can be numerically found using the \texttt{ht} and \texttt{ci} functions in \textsf{R} package \texttt{sbim}.

There is a bias-variance trade-off in the choice of the simulation points $\{\theta_m; m\in 1\cl M\}$.
If the simulation points are chosen within a narrow range, the quadratic approximation to $\mu(\theta)$ will be relatively accurate in that range, resulting in a smaller inference bias.
However, the regression estimates will exhibit greater variance, leading to wider confidence intervals.
In practice, a balance can be achieved by considering both the statistical significance of the third-order term in the Taylor expansion approximation of $\mu(\theta)$ and the size of the constructed confidence region, as demonstrated in Section~\ref{sec:numerical}.


\section{Numerical results}\label{sec:numerical}
We numerically test the simulation-based parameter inference methods developed in Sections~\ref{sec:est_mesle_K1} and \ref{sec:est_proxy}.
Hypothesis tests and construction of confidence intervals were carried out using the \texttt{ht} and \texttt{ci} functions in \textsf{R} package \texttt{sbim} (available at \url{https://CRAN.R-project.org/package=sbim}).
Additional numerical results are provided in Supplementary Section~\ref{sec:additional_num}.

\subsection{Gamma process with Poisson observations}\label{sec:num_gamma_poisson}
We first consider independent, gamma distributed draws $X_{1:n} \overset{iid}{\sim} \Gamma(\gamma, \lambda)$ and conditionally independent Poisson observations $Y_i$ of $X_i$, for $i\in 1\cl n$.
We consider estimating $\lambda$, assuming that $\gamma$ is known.
We generate $n=1000$ observations $y_{1:n}$ for $\gamma=1$, $\lambda=1$.
Simulations $X_{1:n} \overset{iid}{\sim} \Gamma(1, \lambda)$ are carried out at $\lambda = 1.0 \pm 0.001\times k$, $k=0, \dots, 200$ ($M=401$).
Example~\ref{ex:proxy_gamma_poisson} in the supplementary text shows that the MESLE is given by $(n\gamma)/\sum_{i=1}^n y_i$ and that the simulation-based proxy $\lambda_*$ is equal to the true parameter value $\lambda=1$.
Additionally, it shows that $K_1(\lambda) = 2$ and $K_2(\lambda)=1$ are different, and both differ from the Fisher information $\mathcal I(\lambda) = \tfrac{1}{2}$.

\begin{figure}[tp]
\begin{knitrout}
\definecolor{shadecolor}{rgb}{0.969, 0.969, 0.969}\color{fgcolor}

{\centering \includegraphics[width=\maxwidth]{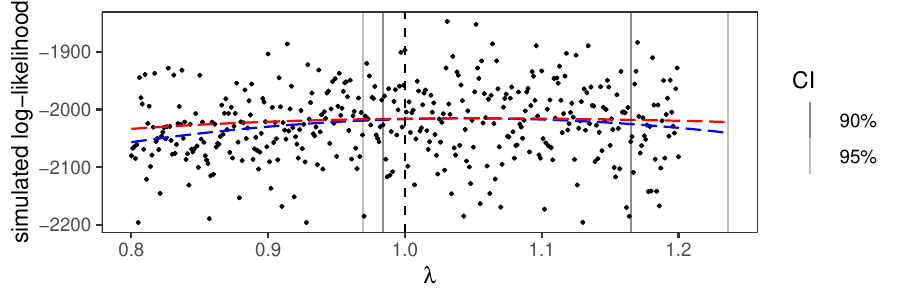} 

}

\end{knitrout}
\caption{Simulated log-likelihoods for the gamma-Poisson process. The 90\% and 95\% confidence intervals constructed for the simulation-based proxy $\lambda_*$ are marked by gray vertical lines. The true value of $\lambda$ is marked by the vertical dashed line. The blue dashed curve indicates the fitted quadratic polynomial, and the red dashed curve the exact log-likelihood function, with a vertical shift for better visual comparison. 
}
\label{fig:gamma_poisson_simll}
\end{figure}

Figure~\ref{fig:gamma_poisson_simll} shows the simulated log-likelihoods and 90\% and 95\% confidence intervals constructed for the simulation-based proxy $\lambda_*$, marked by pairs of gray vertical lines.
The fitted quadratic polynomial is indicated by the blue curve, and the exact log-likelihood by the red curve with a vertical shift for easier comparison with the fitted polynomial.
These curves show that the second order derivative of the fitted quadratic function is different than that of the log-likelihood function, aligning with the fact that $K_2(\lambda)=1$ is not equal to the Fisher information $\mathcal I(\lambda) = \tfrac{1}{2}$.
Hence, using the approximate method by Ionides et al.\@ \cite{ionides2017monte} would lead to misquantified parameter uncertainty.

\begin{figure}[tp]
\begin{knitrout}
\definecolor{shadecolor}{rgb}{0.969, 0.969, 0.969}\color{fgcolor}

{\centering \includegraphics[width=\maxwidth]{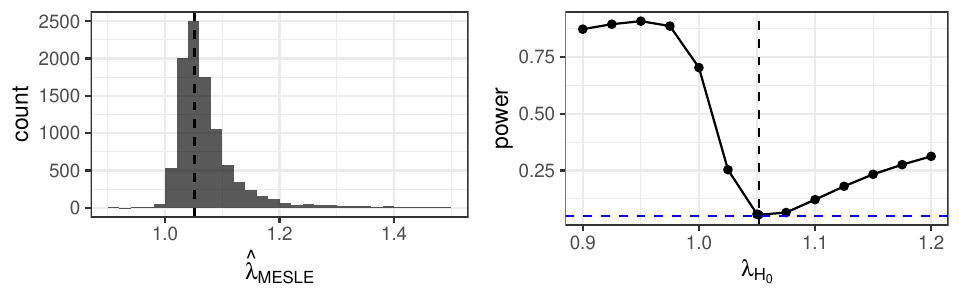} 

}

\end{knitrout}
\caption{The left panel shows the distribution of the estimate $\hat \lambda_\MESLE$ over 10000 replications of simulation-based estimation. The exact value of $\lambda_\MESLE$ is marked by the vertical dashed line. A small number of $\hat\lambda_\MESLE$ that were outside the displayed range were omitted from the plot. 
The right plot shows the estimated rejection probabilities for $H_0: \lambda_\MESLE = \lambda_{H_0}$ at a 5\% significance level for varied null values $\lambda_{H_0}$. The blue horizontal line indicates the significance level. }
\label{fig:gampoi_MESLE_power}
\end{figure}

For the given data $y_{1:n}$, we estimated $\lambda_\MESLE$ by simulating $X_{1:n}$ at varied $\lambda$ values and applying the method introduced in Section~\ref{sec:est_mesle}.
We replicated this experiment ten thousand times.
For each replication, a hypothesis test for the MESLE, $H_0: \lambda_\MESLE = \lambda_{H_0}$, $H_1: \lambda_\MESLE \neq \lambda_{H_0}$, was carried out.
The left panel of Figure~\ref{fig:gampoi_MESLE_power} shows the distribution of the point estimates for $\lambda_\MESLE$.
These estimates are centered around the exact value with an interquartile range of approximately 0.05.
The right panel in Figure~\ref{fig:gampoi_MESLE_power} shows the proportion of the hypothesis tests where the null hypothesis for varied $\lambda_{H_0}$ is rejected at a 5\% significance level.
This plot shows that the empirical significance level aligns with the nominal level and that the test has reasonably high power when the difference between $\lambda_\MESLE$ and the null value $\lambda_{H_0}$ is greater than approximately 0.1.


\begin{figure}[tp]
\begin{knitrout}
\definecolor{shadecolor}{rgb}{0.969, 0.969, 0.969}\color{fgcolor}

{\centering \includegraphics[width=\maxwidth]{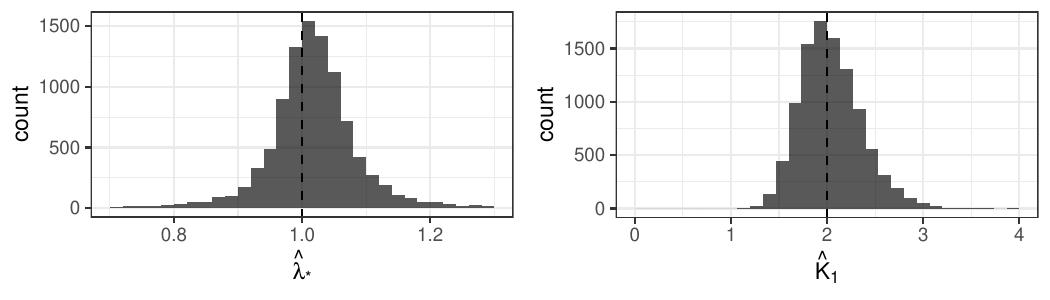} 

}

\end{knitrout}
\caption{The distribution of the estimates for the simulation-based proxy $\lambda_*$ (left) and the estimates for $K_1$ (right). The dashed lines indicate the true values of $\lambda$ and $K_1$.}
\label{fig:gampoi_K1_proxy_dist}
\end{figure}

\begin{figure}[tp]
\begin{knitrout}
\definecolor{shadecolor}{rgb}{0.969, 0.969, 0.969}\color{fgcolor}

{\centering \includegraphics[width=\maxwidth]{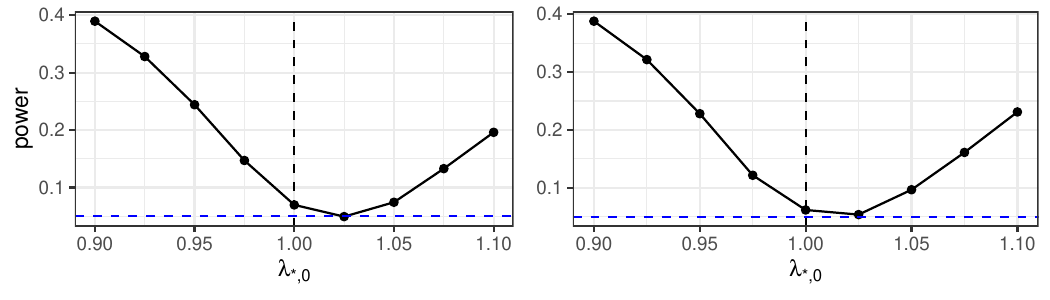} 

}

\end{knitrout}
\caption{The left panel shows the probability of rejecting the null hypothesis $H_0: \lambda_* = \lambda_{*,0}$ at a 5\% significance level for varied null values $\lambda_{*,0}$. The right panel shows the probability of rejecting the null hypothesis when the exact value of $K_1(\lambda_0)$ is used instead of an estimated value.}
\label{fig:gampoi_proxy_power}
\end{figure}


\begin{table}[tp]
  \centering
  \begin{tabular}{lccc}\toprule
    & \multicolumn{3}{c}{Nominal confidence level (\%)}\\\cmidrule{2-4}
    method & 80 & 90 & 95 \\\midrule
    metamodel-based & 77.6 (0.8) & 87.8 (0.7) & 93.2 (0.5) \\
    Ionides et al.\@ \cite{ionides2017monte} & 45.3 (1.0) & 54.9 (1.0) & 63.0 (1.0) \\\bottomrule
  \end{tabular}
  \caption{Percentages of confidence intervals encompassing the true parameter $\lambda_0=1$ using our metamodel-based method (first row) and the method proposed by Ionides et al.\@ \cite{ionides2017monte} (second row) in ten thousand replications. The numbers in parentheses represent twice the standard errors.}
  \label{tab:gampoi_coverage}
\end{table}

We independently generated data $y_{1:n}$ ten thousand times at $\lambda_0=1$ and estimated the simulation based proxy $\lambda_*$ using the method developed in Section~\ref{sec:est_proxy}.
Tests on the simulation-based proxy, $H_0: \lambda_* = \lambda_{*,0}$, $H_1: \lambda_* \neq \lambda_{*,0}$ were carried out for varied null values $\lambda_{*,0}$.
The left panel of Figure~\ref{fig:gampoi_K1_proxy_dist} shows the distribution of the point estimates $\lambda_*$, which was approximately centered at the true parameter value.
The right panel of Figure~\ref{fig:gampoi_K1_proxy_dist} shows the distribution of the estimates for $K_1(\lambda_0)$, estimated using Algorithm~\ref{alg:K1est_block}.
The distribution of $\hat K_1$ is centered at the exact value of $K_1(\lambda_0)=2$.

The left panel in Figure~\ref{fig:gampoi_proxy_power} shows the empirical proportions where $H_0: \lambda_* = \lambda_{*,0}$ is rejected at a 5\% significance level for varied $\lambda_{*,0}$.
The empirical rejection probability was 0.07 at the true null value $\lambda_{*,0}=1$.
The rejection probability was closest to the significance level 0.05 approximately at $\lambda_{*,0} = 1.025$.
One of the reasons for this bias is the variability in the estimated values of $K_1(\lambda_0)$ obtained by Algorithm~\ref{alg:K1est_block}.
If we instead use the exact value $K_1(\lambda_0)$ for the hypothesis tests, the power curve has a smaller bias, as shown by the right plot of Figure~\ref{fig:gampoi_proxy_power}.
However, there is still a bias when the exact $K_1(\lambda_0)$ is used; this is possibly due to the metamodel not being exact or the violation of the assumption that $S_n(Y_{1:n})$ and $\sigma^2(Y_{1:n})$ are independent.

We constructed confidence intervals for $\lambda_*$ using Eq.\@ \eqref{eqn:ci_sbs} ten thousand times.
For comparison, we constructed Monte Carlo confidence intervals using the method of Ionides et al.~\cite{ionides2017monte}, utilizing the \texttt{mcap} function from the \texttt{pomp} package~\cite{king2016statistical, king2023pomp}.
Table~\ref{tab:gampoi_coverage} reports the probabilities that the 80\%, 90\%, and 95\% confidence intervals constructed by both methods encompass the true parameter value $\lambda_0 = 1$.
These results demonstrate that our method constructs confidence intervals that are significantly more accurate than those produced by the method of Ionides et al.~\cite{ionides2017monte}.
The latter exhibits bias because the curvature of the quadratic polynomial fitted to the simulated log-likelihoods generally differs from the Fisher information.


\begin{figure}[tp]
\begin{knitrout}
\definecolor{shadecolor}{rgb}{0.969, 0.969, 0.969}\color{fgcolor}

{\centering \includegraphics[width=\maxwidth]{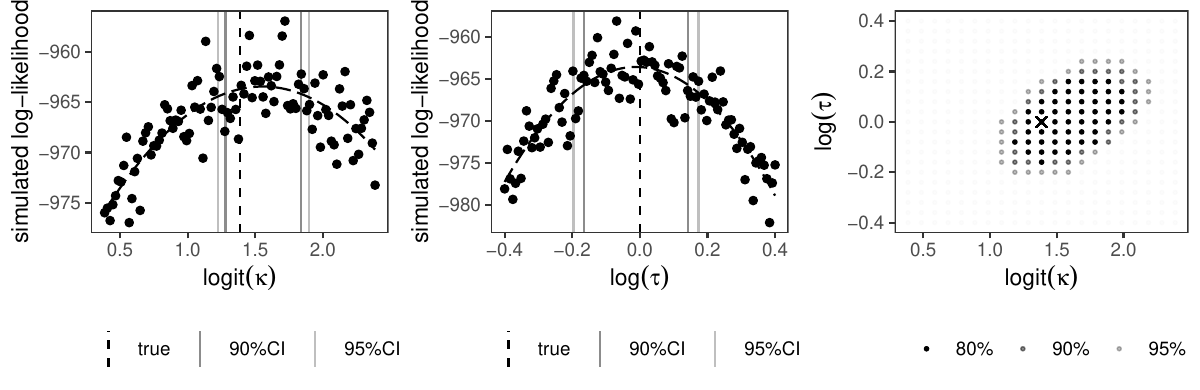} 

}

\end{knitrout}
\caption{Left: Simulated log-likelihoods and constructed confidence intervals for $\text{logit}(\kappa)$ in the stochastic volatility model. Middle: Simulated log-likelihoods and confidence intervals for $\log(\tau)$. Right: Constructed confidence regions for $(\kappa, \tau)$ at 95\%, 90\%, and 80\% confidence levels. The true parameter values are marked by with an `X'.}
\label{fig:stovol_confreg}
\end{figure}

\subsection{Stochastic volatility model}\label{sec:stovol}
We consider a stochastic volatility model, where the distribution of the log rate of return $r_i$ of a stock at time $i$ is described by
\[
  r_i = e^{s_i} W_i, \quad W_i \overset{iid}{\sim} t_5,
\]
where $s_i$ denotes the volatility at time $i$ and $t_5$ the $t$ distribution with five degrees of freedom.
The distribution of the stochastic volatility process $\{s_i\}$ is described by
\[
  s_i = \kappa s_{i-1} + \tau \sqrt{1-\kappa^2} V_i ~~\text{for}~ i>1, \quad s_1 = \tau V_1, \quad   \quad V_i \overset{iid}{\sim} \N(0,1).
\]
The rates of return $r_i$ are observed for $i\in 1\cl n$ where $n=500$. 
We simulate the stochastic volatility process for $\kappa=0.8, \tau=1$ and generate an observed data sequence $r_{1:n}$.

The bootstrap particle filter was run at varied parameter values $\theta = (\kappa, \tau)$ to obtain likelihood estimates using the \textsf{R} package \texttt{pomp} \cite{king2016statistical, king2023pomp}.
Figure~\ref{fig:stovol_confreg} shows the logarithm of the likelihood estimates using one hundred particles.
The left and the middle plots respectively show the confidence intervals for $\text{logit}(\kappa)$ and $\log(\tau)$ where the other parameter was fixed at its true value.
The right plot shows the 80\%, 90\%, and 95\% confidence regions constructed by carrying out the hypothesis tests jointly for both parameters, $H_0: (\kappa_*, \tau_*) = (\kappa_{*,0}, \tau_{*,0})$, $H_1: (\kappa_*, \tau_*) = (\kappa_{*,0}, \tau_{*,0})$, for varied null value pairs and marking those for which the p-value is greater than 20\%, 10\%, and 5\%, respectively.
All three constructed confidence regions encompass the true parameter value.

\begin{figure}[tp]
\begin{knitrout}
\definecolor{shadecolor}{rgb}{0.969, 0.969, 0.969}\color{fgcolor}

{\centering \includegraphics[width=\maxwidth]{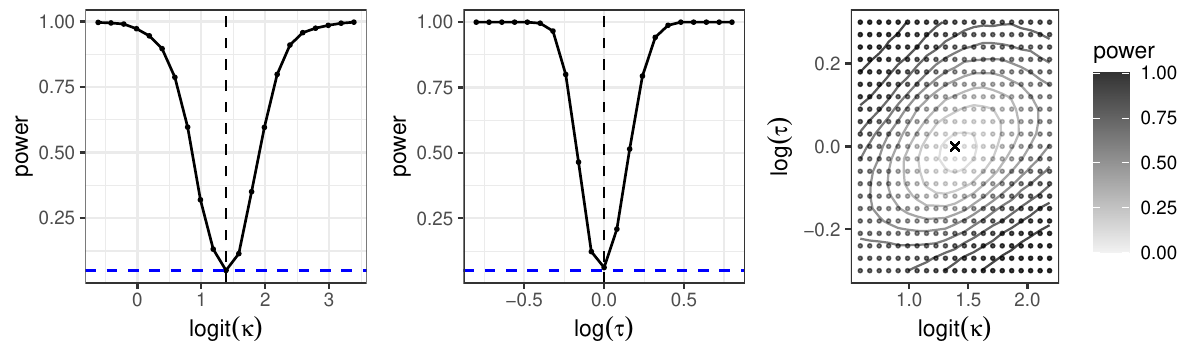} 

}

\end{knitrout}
\caption{Left: Probability of rejecting the null hypothesis $H_0: \kappa_* = \kappa_{*,0}$ for varied null values. Middle: Probabilies of rejecting $H_0: \tau_* = \tau_{*,0}$ for varied null values. Right: Probabilities of rejecting $H_0: (\kappa_*, \tau_*) = (\kappa_{*,0}, \tau_{*,0})$ for varied null value pairs, with level sets indicated. The true parameter values are marked by an `X'. All tests were carried out using a 5\% significance level.}
\label{fig:stovol_power_plots}
\end{figure}

\begin{table}[tp]
  \centering
  \begin{tabular}{cclccc}\toprule
    & & & \multicolumn{3}{c}{Nominal confidence level (\%)}\\\cmidrule{4-6}
    parameter & \# particles & method & 80 & 90 & 95 \\\midrule
    \multirow{4}{*}{$\kappa$} & \multirow{2}{*}{10} & metamodel-based & 79.3 (2.6) & 90.4 (1.9) & 95.4 (1.3) \\
    & & Ionides et al.\@ \cite{ionides2017monte} & 63.1 (3.1) & 76.2 (2.7) & 84.8 (2.3) \\\cmidrule{2-6}
    & \multirow{2}{*}{100} & metamodel-based & 79.2 (2.6) & 89.3 (2.0) & 94.7 (1.4) \\
    & & Ionides et al.\@ \cite{ionides2017monte} & 76.7 (2.7) & 88.2 (2.0) & 94.3 (1.5) \\\midrule
    \multirow{4}{*}{$\tau$} & \multirow{2}{*}{10} & metamodel-based & 76.6 (2.7) & 86.5 (2.2) & 92.4 (1.7) \\
    & & Ionides et al.\@ \cite{ionides2017monte} & 60.5 (3.1) & 75.5 (2.7) & 83.5 (2.3) \\\cmidrule{2-6}
    & \multirow{2}{*}{100} & metamodel-based & 77.3 (2.6) & 90.1 (1.9) & 94.9 (1.4) \\
    & & Ionides et al.\@ \cite{ionides2017monte} & 76.8 (2.7) & 88.5 (2.0) & 94.6 (1.4) \\\bottomrule
  \end{tabular}
  \caption{Percentages of confidence intervals encompassing the true parameter values for $\kappa$ and $\tau$ using our metamodel-based method (first row) and the method of Ionides et al.\@ \cite{ionides2017monte} (second row). The confidence intervals are constructed one thousand times by running the particle filter with ten or one hundred particles. The numbers in parentheses represent twice the standard errors.}
  \label{tab:stovol_coverage}
\end{table}

Hypothesis tests were replicated one thousand times, each time generating a new observation sequence under $\kappa=0.8$ and $\tau=1$ and estimating the likelihood using one hundred particles.
Figure~\ref{fig:stovol_power_plots} shows the estimated probabilities of rejecting the null hypothesis at a 5\% significance level for varied null values.
For tests on either parameter, the empirically estimated significance levels were close to the nominal significance level of 0.05, and the power increased as the null values diverged from the true values.
The right plot of Figure~\ref{fig:stovol_power_plots} shows the rejection probability for the simultaneous test on both parameters.
The empirical significance level was approximately 7.3\%, and the rejection probability increased as $(\kappa_{*,0}, \tau_{*,0})$ moved farther from the true values.

Table~\ref{tab:stovol_coverage} shows the percentages of the confidence intervals constructed for $\kappa$ and $\tau$ that encompass the true values.
The likelihood estimates were obtained using either ten or one hundred particles.
The confidence intervals constructed using our method include the true parameter values with probabilities close to the nominal confidence levels in both cases.
However, the confidence intervals constructed using the method proposed by Ionides et al.\@ \cite{ionides2017monte} exhibit significantly lower coverage probabilities when a small number of particles are used for likelihood estimation.
As the Monte Carlo variance of the log-likelihood estimate increases with fewer particles, the fitted quadratic polynomial may deviate significantly from the exact log-likelihood function.
This leads to a misquantification of parameter uncertainty in the method of Ionides et al.\@ \cite{ionides2017monte}.
In contrast, our method correctly distinguishes bewteen $K_1$ and $K_2$ in the LAN of the simulated log-likelihoods (Proposition~\ref{prop:lan_sim}), maintaining accuracy even when weak or imprecise Monte Carlo likelihood estimators are used.

\subsection{Stochastic SEIR model for population dynamics of measles transmission}\label{sec:measles}
We demonstrate our parameter inference procedure applied to a mechanistic model describing the population dynamics of measles transmission in England and Wales between 1950 and 1964.
Weekly reported case data for twenty cities were analyzed by He et al.\@ \cite{he2009plug} using a stochastic compartment model consisting of the susceptible (S), exposed (E), infectious (I), and recovered (R) compartments.
Partial observations of the compartment sizes are given by weekly reported case numbers, which are random fractions of weekly aggregate transitions from the infectious to the recovered compartment. 
We carried out parameter inference for the basic reproduction number $R_0$.
Model details as well as additional parameter inference results are given in the supplementary text Section~\ref{sec:supp_measles}.

\begin{figure}[t]
\begin{knitrout}
\definecolor{shadecolor}{rgb}{0.969, 0.969, 0.969}\color{fgcolor}

{\centering \includegraphics[width=\maxwidth]{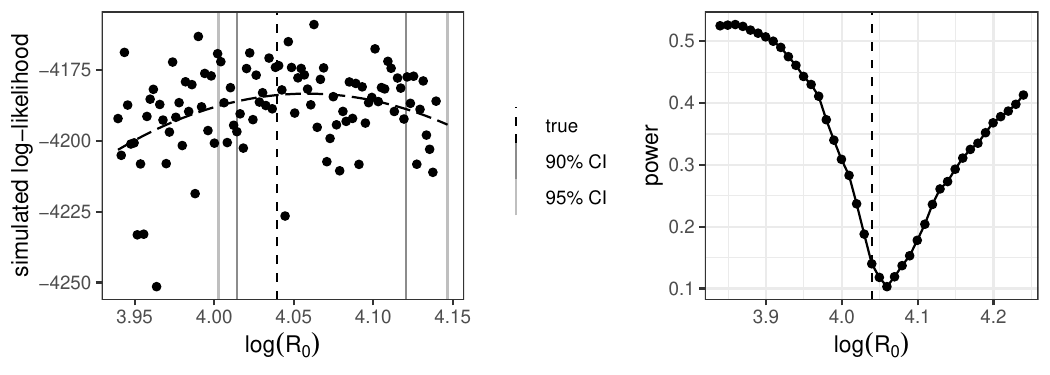} 

}

\end{knitrout}
\caption{Left: Simulated log-likelihoods and the constructed 90\% and 95\% confidence intervals for $\log(R_0)$. Right: Probability of rejecting the null hypothesis for varied null values of $R_0$ at a 5\% significance level. The true value is indicated by the dashed vertical line.} 
\label{fig:measles_R0}
\end{figure}

We simulated the SEIR model at a suitably chosen parameter vector and generated a sequence of weekly reported cases data.
Unbiased likelihood estimate for the observed data sequence were obtained for varied parameters using the bootstrap particle filter via the \textsf{R} package \texttt{pomp}.
The left plot of Figure~\ref{fig:measles_R0} shows the simulated log-likelihoods for varied $\log(R_0)$ and the constructed 90\% and 95\% confidence intervals.
Simulations were carried out at $M=100$ points uniformly placed between the exact value $\pm 0.1$ on the log scale.
We replicated simulation-based inference for $R_0$ one thousand times.
The right plot of Figure~\ref{fig:measles_R0} shows the probability of rejecting the null hypothesis at a 5\% significance level for varied null values of $R_0$.
The empirical significance level was somewhat higher than the nominal significance level.
However, the rejection probability was minimized near the true parameter value, indicating a reasonably small bias in parameter inference.

\section{Comparison with pseudo-marginal MCMC}\label{sec:comp_pmcmc}

\begin{table}[tp]
  \centering
  \begin{tabular}{lcccccc}\toprule
    & \multicolumn{6}{c}{Number of simulations}\\\cmidrule{2-7}
    method & 100 & 1000 & 3000 & \ensuremath{10^{4}} & \ensuremath{3\times 10^{4}} & \ensuremath{10^{5}}\\\midrule
    pseudo-marginal MCMC & 0.17 & 0.21 & 0.25 & 0.30 & 0.36 & 0.40\\
    metamodel-based & 0.54 &  18 &  58 & 190 & 600 & 1800 \\\bottomrule
  \end{tabular}
  \caption{Effective sample sizes for pseudo-marginal MCMC and corresponding efficiency measures for our metamodel-based method, with varying numbers of simulations.}
  \label{tab:pmcmc_sbi_ess}
\end{table}
\begin{figure}[tp]
\begin{knitrout}
\definecolor{shadecolor}{rgb}{0.969, 0.969, 0.969}\color{fgcolor}

{\centering \includegraphics[width=\maxwidth]{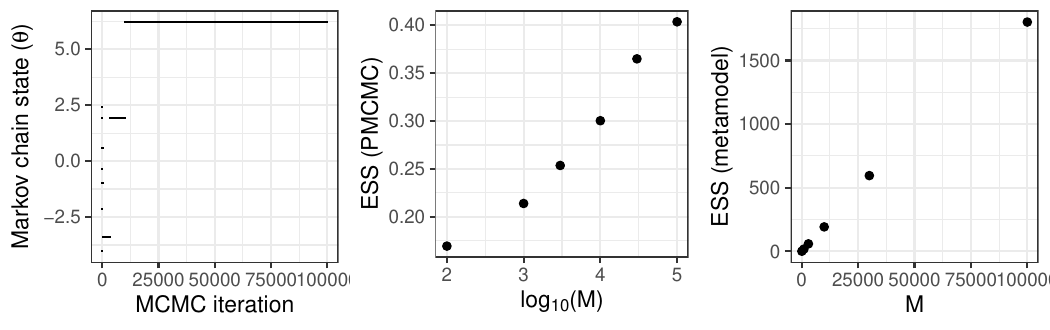} 

}

\end{knitrout}
\caption{Left: Trace plot of a pseudo-marginal MCMC run with length $M=10^5$ for Example~\ref{ex:comp_pmcmc}. Middle: The effective sample size of pseudo-marginal MCMC scales approximately logarithmically with the length of the Markov chain. Right: The corresponding efficiency measure for our metamodel-based method scales linearly with the number of simulations.}
\label{fig:pmcmc_sbi_ess}
\end{figure}

We compare our metamodel-based inference method to a pseudo-marginal Markov chain Monte Carlo method in terms of sampling efficiency.
For a metamodel-based method, the condition for parameter identifiability \eqref{eqn:sig_noise_ratio},
\[
  (\delta \theta)^\top c(y_{1:n}) (\delta \theta) \gtrsim \frac{\sigma(y_{1:n})}{\sqrt M}
\]
implies that for given data $y_{1:n}$, the standard error of our metamodel-based parameter estimator scales approximately as $\Vert \delta \theta \Vert = \O(M^{-1/4})$ as the number of simulation $M$ increases.
We will demonstrate that pseudo-marginal MCMC methods scale much more poorly, with the rate of $\Vert \delta \theta \Vert$ scaling as $\O((\log M)^{-1/4})$, which is significantly worse than our metamodel-based method.

Pseudo-marginal Markov chain Monte Carlo enables Bayesian inference where an unbiased estimator $\hat L(\theta; y_{1:n})$ of the likelihood $L(\theta;y_{1:n})$ is available \cite{andrieu2009pseudo}.
When the unbiased likelihood estimator is obtained for a POMP model using the particle filter, the method is referred to as particle Markov chain Monte Carlo (PMCMC) \cite{andrieu2010particle}.
Pseudo-marginal MCMC constructs a Markov chain having the posterior density 
\[
  \pi(\theta|y_{1:n}) \propto h(\theta) \cdot L(\theta;y_{1:n})
\]
as a stationary distribution, where $h(\theta)$ denotes the density of a prior distribution.
Suppose that at a certain point, the current state of the constructed Markov chain is $\theta$ and an unbiased likelihood estimate $\hat L(\theta; y_{1:n})$ has been obtained.
A candidate $\theta'$ for the next state of the Markov chain is proposed using a proposal kernel with density $q(\theta'|\theta)$.
If we denote by $\hat L(\theta' ; y_{1:n})$ a new Monte Carlo likelihood estimate obtained by running simulations under $\theta'$, the proposed candidate $\theta'$ is accepted with probability
\begin{equation}
  \min\left( 1, \frac{h(\theta')\hat L(\theta';y_{1:n}) q(\theta|\theta') }{h(\theta)\hat L(\theta; y_{1:n}) q(\theta'|\theta)} \right).
  \label{eqn:pmcmc}
\end{equation}
The elements of the constructed Markov chain are considered as approximate draws from the target posterior distribution.

We numerically compare pseudo-marginal MCMC with our method using the following example.

\begin{ex}\label{ex:comp_pmcmc}
  Consider $X_{1:n} \overset{iid}{\sim} \N(\theta, \tau^2)$ and conditionally independent observations $Y_i|X_i \sim \N(X_i, 1)$, $i\in 1\cl n$, where $\tau = 30$ is known.
  Suppose that observations $y_{1:n}$ are available with $n=200$.
  Independent simulations $X_{1:n}$ are drawn from the $\N(\theta, \tau^2)$ distribution, and the simulated log-likelihood is given by $\ell^S(\theta) = -\frac{1}{2} \sum_{i=1}^n (X_i - y_i)^2 -\frac{n}{2} \log (2\pi)$.
  The exact MLE is given by the sample mean, $\bar y = \frac{1}{n} \sum_{i=1}^n y_i$.
  Marginally, $Y_i \overset{iid}\sim \N(\theta, \tau^2+1)$, and the exact 95\% confidence interval for $\theta$ is given by $\bar y \pm z_{0.025} \cdot \sqrt{(\tau^2+1)/n}$ where $P[\N(0,1) > z_{0.025} \approx 1.96] = 0.025$.
  For Bayesian inference, we consider a flat prior $h(\theta) \equiv 1$.
  The posterior distribution is given by $\theta | y_{1:n} \sim \N(\bar y, \frac{\tau^2+1}{n})$, and the 95\% Bayesian credible interval is identical to the 95\% confidence interval, $\bar y \pm z_{0.025} \cdot \sqrt{(\tau^2+1)/n}$.
  In this section, pseudo-marginal MCMC was run with proposal kernel $\theta' \sim \N(\theta, 3^2)$.
\end{ex}

Consider a Markov chain of length $M$ constructed by pseudo-marginal MCMC.
This chain uses the same number of simulations as a metamodel-based method that employs $M$ simulated log-likelihoods.
In a pseudo-marginal MCMC, a proposed value $\theta$ is accepted if the associated simulated log-likelihood $\ell^S(\theta)$ is relatively large compared to the previous values in the chain.
However, if the data size $n$ is large, the Monte Carlo standard deviation $\sigma$ of the simulated log-likelihoods scales as $\O(\sqrt n)$ and thus dominates the difference in the log-likelihood function, which remains $\O(1)$ over the range of the exact Bayesian credible interval.
Thus, whether a proposed value is accepted is almost entirely dependent on chance variation in $\ell^S(\theta)$, and the probability that the Markov chain changes value at the $m$-th step is approximately $m^{-1}$.
Hence, the total number of updates in a chain of length $M$ scales as $\sum_{m=1}^M m^{-1} \approx \log M$.
The left plot of Figure~\ref{fig:pmcmc_sbi_ess} shows a trace plot of a pseudo-marginal MCMC chain of length $M=10^5$ for Example~\ref{ex:comp_pmcmc}.
This chain remains at the same value for approximately 90\% of its length.
The MCMC parameter estimate $\hat\theta^M_{MCMC}$, obtained by averaging the values in the chain, has variability comparable to that of a single draw from the posterior.
The efficiency of MCMC can be assessed using the effective sample size (ESS), defined as
\begin{equation}
  \text{ESS} = \frac{\text{Var}_\pi (\theta)}{\text{Var}(\hat \theta^M_{MCMC})},
  \label{eqn:ESS}
\end{equation}
where the numerator represents the variance of the target posterior distribution for $\theta$, and the denominator represents the Monte Carlo variance of the MCMC estimator $\hat \theta^M_{MCMC}$.

We estimated the ESS for pseudo-marginal MCMC by computing the Monte Carlo variance $\text{Var}(\hat \theta^M_{MCMC})$ using one thousand independently constructed Markov chains.
For each chain, MCMC estimates for $\theta$ were obtained by taking the running average at steps $M=100, 1000, 3000, 10^4, 3\times 10^4$, and $10^5$ after discarding the first one hudred iterations as burn-in.
The effective sample sizes were then calculated by substituting the sample variance of these one thousand MCMC estimates into the denominator of \eqref{eqn:ESS}.
The exact value of the numerator was given by $(\tau^2+1)/n$.
For comparison, we estimated the parameter $\theta$ using our metamodel-based method with the same range of simulation counts $M$.
A corresponding efficiency measure was computed using Equation~\ref{eqn:ESS}, replacing the denominator with the sample variance of the parameter estimates across one thousand replications.
Table~\ref{tab:pmcmc_sbi_ess} presents the ESS values for both methods.
For pseudo-marginal MCMC, the effective sample size grew very slowly and remained extremely low even at $M=10^5$.
The fact that the ESS was less than one suggests that the chains had not yet effectively explored the target posterior distribution and that the MCMC estimate $\hat \theta^M_{MCMC}$ was essentially determined by a single best simulation outcome, similar to what was demonstrated in the top panel of Figure~\ref{fig:intro_OU}.
The middle plot of Figure~\ref{fig:pmcmc_sbi_ess} shows that these ESS values scaled logarithmically with the chain length $M$. 
In contrast, the equivalent ESS values for our metamodel-based method were dramatically higher than those for pseudo-marginal MCMC.
The right plot of Figure~\ref{fig:pmcmc_sbi_ess} shows that the ESS scaled linearly with the number of simulations $M$, indicating no diminishing returns.

\begin{figure}[tp]
\begin{knitrout}
\definecolor{shadecolor}{rgb}{0.969, 0.969, 0.969}\color{fgcolor}

{\centering \includegraphics[width=\maxwidth]{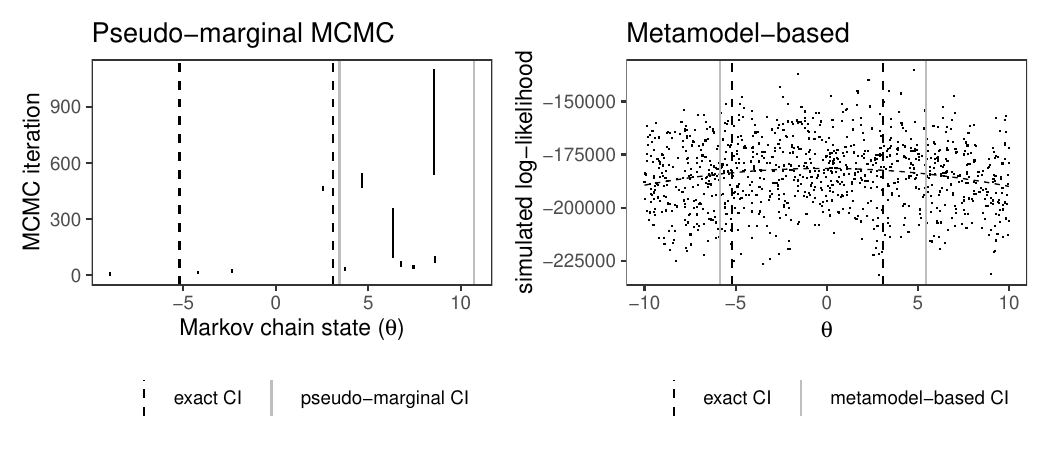} 

}

\end{knitrout}
\caption{Left: Trace plot of a Markov chain of length 1100 using pseudo-marginal MCMC for Example~\ref{ex:comp_pmcmc}.
  The exact 95\% credible interval is marked by two vertical dashed lines.
  The 95\% credible interval constructed from the MCMC draws, after discarding the first one hundred as burn-in, is marked by vertical gray lines.
  Right: A 95\% confidence interval for $\theta$ generated using our metamodel-based method with $M=1000$ simulations, marked by gray vertical lines.
  The exact 95\% confidence interval, indicated by dashed vertical lines, is identical to the exact 95\% credible interval in the left panel.
}
\label{fig:pmcmc_sbi_demonstration}
\end{figure}

\begin{figure}[tp]
\begin{knitrout}
\definecolor{shadecolor}{rgb}{0.969, 0.969, 0.969}\color{fgcolor}

{\centering \includegraphics[width=\maxwidth]{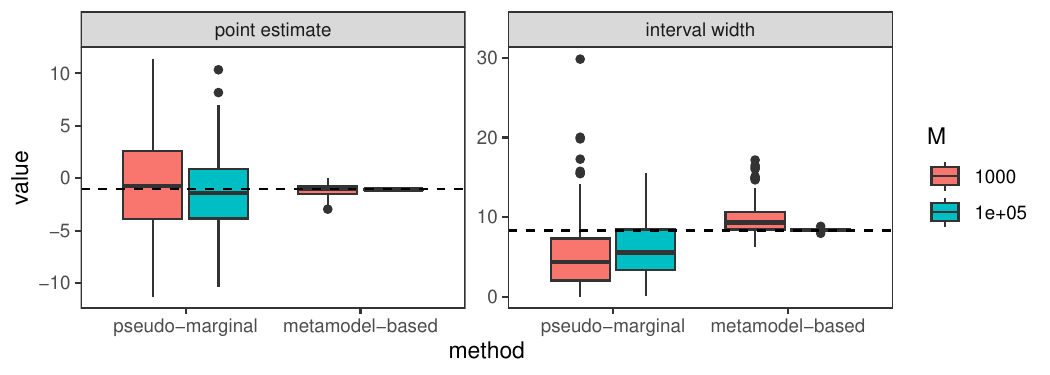} 

}

\end{knitrout}
\caption{Left: Distribution of point estimates for $\theta$ obtained by pseudo-marginal MCMC and our metamodel-based method using $M=10^3$ and $10^5$ simulations.
  The exact MLE ($=\bar y$) is indicated by the horizontal line.
  Right: Distribution of the widths of the constructed 95\% credible intervals for pseudo-marginal MCMC and the 95\% confidence intervals constructed using our metamodel-based method.
  The width of the exact 95\% credible and confidence intervals is indicated by the horizontal line.}
\label{fig:pmcmc_sbi_boxplots}
\end{figure}

Next, we constructed Monte Carlo interval estimates for $\theta$ using both methods.
The left plot of Figure~\ref{fig:pmcmc_sbi_demonstration} shows a 95\% credible interval constructed using $M=1000$ pseudo-marginal MCMC draws, after discarding the first one hundred iterations as burn-in.
This interval estimate exhibits a substantial bias relative to the exact credible interval, as the constructed chain remains outside the exact 95\% credible interval for most of its run.
In contrast, the right panel of Figure~\ref{fig:pmcmc_sbi_demonstration} shows that a 95\% confidence interval constructed using our metamodel-based method is reasonably close to the exact confidence interval, with a moderately greater width due to simulation randomness.

Figure~\ref{fig:pmcmc_sbi_boxplots} shows the distribution of point estimates and the widths of constructed 95\% credible and confidence intervals for two hundred replications of each method.
The number of simulations were either $M=10^3$ or $10^5$.
The left panel illustrates that the point estimates derived from pseudo-marginal MCMC exhibit significantly higher variation than those obtained by our metamodel-based method.
For pseudo-marginal MCMC, the variation decreases only marginally when the number of iterations $M$ increases from $10^3$ to $10^5$.
In contrast, point estimates obtained by our metamodel-based method have both high precision and high accuracy.
The right panel of Figure~\ref{fig:pmcmc_sbi_boxplots} shows that the widths of the 95\% credible intervals constructed using pseudo-marginal MCMC present substantial variation and a significant downward bias relative to the exact 95\% credible interval, indicating an underestimation of parameter uncertainty.
In contrast, the 95\% confidence intervals constructed using our metamodel-based method show only moderate upward bias and variation when $M=1000$.
With $M=10^5$ simulations, the simulation-based uncertainty is nearly eliminated, yielding Monte Carlo confidence intervals that closely match the exact confidence interval.


The fact that the ESS of pseudo-marginal MCMC scales logarithmically with $M$ implies that a desired level of precision in parameter estimation may not be achieved in practice.
Specifically, in the supplementary text, Section~\ref{sec:supp_pmcmc_scaling}, we argue that parameter estimates obtained by pseudo-marginal MCMC have standard errors approximately given by
\begin{equation}
  \text{standard error of }\hat\theta^M_{MCMC}
  \approx \Vert c \Vert^{-1/2} \sigma^{1/2} (\log M)^{-1/4},
  \label{eqn:pmcmc_scaling}
\end{equation}
where $c$ indicates the curvature of the expected simulated log-likelihood $\mu(\theta)$ and $\sigma$ the standard deviation of the simulated log-likelihoods.
This result is numerically supported by Figure~\ref{fig:pmcmc_sbi_boxplots}, which shows a slight reduction in the variability of parameter estimates from pseudo-marginal MCMC when $M$ increases from $10^3$ to $10^5$.
Hence, to achieve a standard error of $\O(1/\sqrt n)$, one must have
\[
  \log M = \O(\sigma^2),
\]
given that $\Vert c \Vert = \O(n)$.
These results align with the findings of Pitt et al.\@ \cite{pitt2012some} and Doucet et al.\@ \cite{doucet2015efficient}, which suggest that the computational budget is optimally allocated when a sufficiently large amount of Monte Carlo effort is invested to ensure that the Monte Carlo variance of the log-likelihood estimator, $\sigma^2$, is close to one.

When the particle filter is used for likelihood estimation with $J$ particles, B\'erard et al.~\cite{berard2014lognormal} establish in a central limit theorem-type result that $\sigma^2$ asymptotically scales as $\O(n/J)$ uncer certain regularity conditions.
This suggests that choosing $J = \O(n)$ particles may achieve optimal efficiency for pseudo-marginal MCMC.
However, this result has practical implications only when each observation $y_i$, $i\in 1\cl n$, is low dimensional.
The claimed asymptotic normality of $\ell^S$ is not attained unless $J$ is exponentially large in the dimension of $y_i$, due to the highly skewed distribution of the measurement density $g_i(y_i|X_i)$.
The requirement for the number of particles to increase exponentially with the space dimension is well known in the particle filter literature and is often referred to as the curse of dimensionality \cite{bengtsson2008curse, snyder2008obstacles}.
If the dimension $k$ of each $y_i$ is at least moderately large, the distribution of the simulated log-likelihood $\ell^S_i$ can be approximated as
\begin{equation}
  \hat \ell^S_i = \log\left\{\frac{1}{J} \sum_{j=1}^J g_i(y_i|X_i^j)\right\}
  \approx \max_{j\in 1\cl J} \log g_i(y_i|X_i^j) - \log J.
  \label{eqn:sll_max_approx}
\end{equation}
However, the variance of the right-hand side of \eqref{eqn:sll_max_approx} scales as $\O(k/\log J)$, provided that the distribution of $\log g_i(y_i|X_i^j)$ is approximately normal.
See the supplementary text (Section~\ref{sec:supp_sll_max_approx}) for numerical justification.
This suggests that, to achieve $\sigma^2\approx 1$ for optimal efficiency in pseudo-marginal MCMC, $J=\O(e^k)$ particles should be used for likelihood estimation at each MCMC iteration.
In contrast, our metamodel-based inference method enables robust and accurate parameter inference even when $\sigma^2$ is large, as the required number of simulations scales linearly with $\sigma^2$.


\begin{figure}[tp]
\begin{knitrout}
\definecolor{shadecolor}{rgb}{0.969, 0.969, 0.969}\color{fgcolor}

{\centering \includegraphics[width=\maxwidth]{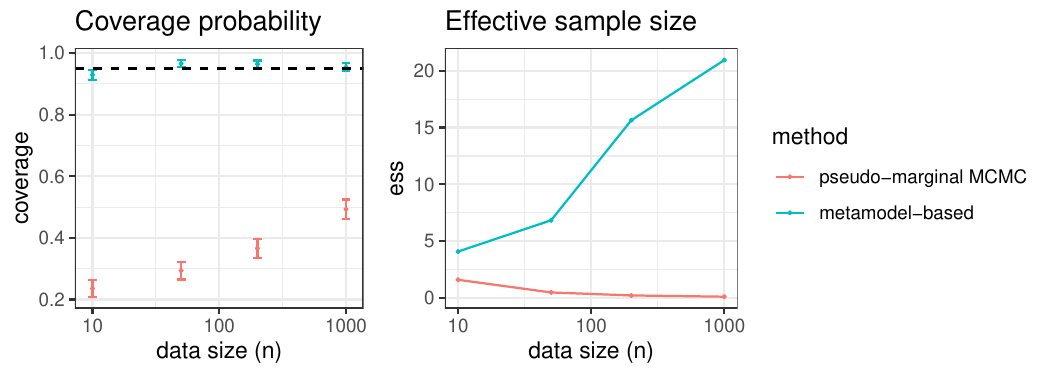} 

}

\end{knitrout}
\caption{Left: Coverage probabilities of the 95\% credible intervals constructed using pseudo-marginal MCMC and the 95\% confidence intervals constructed using our metamodel-based method, with data size ranging from ten to one thousand. The nominal 95\% level is indicated by a horizontal dashed line, and the error bars represent plus or minus two standard errors. Right: Effective sample sizes for pseudo-marginal MCMC and our metamodel-based method, averaged over one hundred generated datasets, for varying data sizes. }
\label{fig:pmcmc_sbi_coverage}
\end{figure}

We compared the coverage probabilities of interval estimates and the effective sample sizes of pseudo-marginal MCMC and our proposed method as the data size $n$ varied among 10, 50, 200, and 1000.
The left plot of Figure~\ref{fig:pmcmc_sbi_coverage} presents the proportion of 95\% credible and confidence intervals that contain the true parameter value, $\theta=0$.
The credible intervals were constructed using $M=1000$ samples from pseudo-marginal MCMC after discarding the first 100 burn-in cycles.
The confidence intervals for our metamodel-based method were constructed using the same number of simulations ($M=1000$).
Approximately 95\% of the intervals constructed by our metamodel-based approach contained the true parameter value.
In contrast, the credible intervals obtained from pseudo-marginal MCMC exhibited notably lower coverage.
The underestimation of parameter uncertainty in pseudo-marginal MCMC is due to its very low effective sample size.

The right plot of Figure~\ref{fig:pmcmc_sbi_coverage} shows the average effective sample sizes (ESS) for pseudo-marginal MCMC and our metamodel-based method across the same range of $n$.
The estimated ESS values were averaged over 100 generated datasets.
For each dataset, the ESS was computed by comparing the sample standard deviation of the parameter estimates with the exact posterior standard deviation using Equation~\ref{eqn:ESS}.
The ESS for pseudo-marginal MCMC decreases as the data size increases due to the increasing variability and skewness in the distribution of the Monte Carlo likelihood estimator.
In contrast, the ESS for the metamodel-based method is significantly higher and increases with $n$, as the improved signal-to-noise ratio of the simulated log-likelihoods enables precise parameter estimation using the metamodel.
Both plots in Figure~\ref{fig:pmcmc_sbi_coverage} demonstrate that the metamodel-based approach scales significantly better with increasing data size than pseudo-marginal MCMC.

\section{Automatic tuning algorithms}\label{sec:autotuning}
In this section, we introduce two automatic tuning algorithms that enhance the applicability of the inference methods developed in this paper.
The first algorithm automatically adjusts the weights assigned to simulated outcomes, ensuring that the quadratic approximation to the simulated log-likelihoods introduces little bias.
Specifically, the weights of the parameter points far from the estimated MESLE are discounted so that the third order term in a cubic regression becomes statistically insignificant.
This automatic weight adjustment scheme can be incorporated into Algorithm~\ref{alg:sbi} to improve hypothesis testing and confidence interval construction.
The second algorithm proposes the next simulation point to achieve near-optimal efficiency.
This point is selected to minimize the Monte Carlo variance of the parameter estimate while ensuring that the quadratic approximation to the simulated log-likelihoods remain valid.
Both of these algorithms are implemented in the \textsf{R} package \texttt{sbim}.
Further details on these algorithms are provided in the supplementary text (Section~\ref{sec:supp_autotuning}.)

\subsection{Automatic weight adjustments for bias reduction}\label{sec:autoAdjust}
\begin{algorithm}[t]
  \caption{Automatic weight adjustments for bias reduction}
  \label{alg:autoAdjust}
  \begin{algorithmic}[1]
    \State Fit a quadratic polynomial to $(\theta_m,\, \ell^S(\theta_m))$ with weights $w_m$, $m \in 1\cl M$, to obtain a first-stage quadratic approximation $q_2(\theta) = \hat a + \hat b^\top \theta + \theta^\top \hat c \theta$
    \State Let $\hat \theta_\MESLE = -\tfrac{1}{2}\hat c^{-1} \hat b = \arg\max_\theta q_2(\theta)$ be the estimated MESLE
    \State Let $g \gets \infty$
    \Loop
      \State Weight adjustments: $w_m^{adj} = w_m \cdot \exp(-\{q_2(\hat\theta_\MESLE) - q_2(\theta_m)\}/g)$
      \State Update $q_2$ and $\hat\theta_\MESLE$ using the adjusted weights
      \State Fit a cubic polynomial to $(\theta_m,\, \ell^S(\theta_m))$ with weights $w_m^{adj}$, $m \in 1\cl M$
      \State Let $p_\text{cubic}$ be the p-value for the significance of the cubic term
      \If {$p_\text{cubic}<0.01$} \Comment{cubic term is significant, decrease $g$}
        \If {$g=\infty$}
          \State Let $g \gets q_2(\hat\theta_\MESLE) - \min_{m\in 1\cl M} q_2(\theta_m)$
        \Else
          \State Let $g \gets g / 1.8$
        \EndIf
      \ElsIf {$p_\text{cubic}>0.3$} \Comment{cubic term is not significant, increase $g$ for efficiency}
        \State Let $g \gets 1.3 \cdot g$
      \Else
        \State Break from loop
      \EndIf
    \EndLoop
  \end{algorithmic}
\end{algorithm}

Obtaining simulated log-likelihoods across a wide range of parameter values increases estimation efficiency but also introduces bias due to the approximation error of a quadratic polynomial.
We develop an algorithm that automatically balances efficiency and accuracy by adjusting the weights $w_m$ assigned to the parameter value $\theta_m$.
Denoting by $q_2(\theta)$ the quadratic polynomial fitted to $(\theta_m, \ell^S(\theta_m))$ with weights $w_m$, the weight for the $m$-th point is adjusted as
\[
  w_m^{adj} \gets w_m \cdot \exp\left(-\frac{q_2(\hat\theta_\MESLE) - q_2(\theta_m)}{g}\right),
\]
ensuring that points far from the estimated MESLE, $\hat \theta_\MESLE$, have more heavily discounted weights.
The scalar $g$ is tuned using an iterative procedure as follows.
Let $p_\text{cubic}$ be the p-value indicating the significance of the cubic term when a cubic polynomial is fitted to $(\theta_m, \ell^S(\theta_m))$ with the adjusted weights $w_m^{adj}$ for $m\in 1\cl M$.
\begin{itemize}
\item If the cubic term is highly statistically significant (e.g., $p_\text{cubic} < 0.01$), the quadratic approximation $q_2$ is likely introducing significant error.
  In this case, we decrease $g$ to narrow the range of parameter values that effectively contribute to parameter estimation.
\item Conversely, If the cubic term is insignificant (e.g., $p_\text{cubic} > 0.3$), the quadratic approximation $q_2$ is sufficiently accurate, allowing for broader exploration of the parameter space to improve estimation efficiency.
  In this case, we increase $g$.
\end{itemize}
The value of $g$ is iteratively adjusted until $p_\text{cubic}$ suggests a reasonable balance between efficiency and accuracy.
Algorithm~\ref{alg:autoAdjust} summarizes this procedure.
The functions \texttt{ht} and \texttt{ci} in the \texttt{sbim} package include an option to automatically adjust the weights before performing hypothesis testing or constructing confidence intervals.

\subsection{Selecting the next simulation point for near-optimal efficiency}\label{sec:optDesign}
\begin{algorithm}[t]
  \caption{Near-optimal selection of the next simulation point}
  \label{alg:optDesign}
  \begin{algorithmic}[1]
    \State \textbf{Input}: simulation points $\{\theta_m; m\in 1 \cl M\}$; simulated log-likelihoods $\{\ell^S(\theta_m); m\in 1 \cl M\}$; weights $\{w_m; m\in 1\cl M\}$
    \State \textbf{Output}: a proposal for the next simulation point $\theta_{M+1}$
    \State Adjust weights $w_m$ to reduce the bias due to quadratic approximation (Algorithm~\ref{alg:autoAdjust})
    \State Fit a quadratic polynomial to $(\theta_m,\, \ell^S(\theta_m))$ with the adjusted weights $w_m^{adj}$, $m \in 1\cl M$ and let $\hat A = (\hat a, \hat b^\top, \vech(\hat c)^\top)^\top$ be the estimated coefficients.
    \State Define $\mathrm{STV}$ by \eqref{eqn:STV} as a function of $\theta_{M+1}$
    \State Find $\arg\min_{\theta_{M+1}} STV(\theta_{M+1})$ using a gradient-based numerical optimization algorithm such as BFGS and return the minimizer
  \end{algorithmic}
\end{algorithm}

Given the costs associated with simulations, efficiently selecting simulation points is practically important.
In particular, experiments should be designed to minimize the Monte Carlo variability of parameter estimates for a given number of simulations.
We develop an algorithm that sequentially proposes the next simulation point to achieve near-optimal efficiency.

Suppose that currently there are $M$ simulated log-likelihoods available, obtained at $\theta_m$, $m\in 1\cl M$.
Let $\hat A = (\hat a, \hat b^\top, \vech(\hat c)^\top)^\top$ be the coefficients of the fitted quadratic polynomial with weights adjusted as described in Section~\ref{sec:autoAdjust}.
We see from \eqref{eqn:varAhat} that the Monte Carlo variance of $\hat A$ is given by
\[
  \text{Var}(\hat A) = \sigma^2 \{ {\theta_{1:M}^{0:2}}^\top W^{adj} \theta_{1:M}^{0:2} \}^{-1}
\]
where $W^{adj}$ is the diagonal matrix having the adjusted weights $w_m^{adj}$ as its diagonal entries.
Applying the delta method, we can approximate the Monte Carlo variance of the estimated MESLE, $\hat\theta_\MESLE = -\tfrac{1}{2} \hat c^{-1} \hat b$, by
\[
  \Var(\hat \theta_\MESLE) 
  \approx \sigma^2 \left( \frac{\partial \hat \theta_\MESLE}{\partial \hat A} \right) \{ {\theta_{1:M}^{0:2}}^\top W^{adj} \theta_{1:M}^{0:2} \}^{-1} \left( \frac{\partial \hat \theta_\MESLE}{\partial \hat A} \right)^\top.
\]
After a new simulation is performed at $\theta_{M+1}$, the variance of $\hat\theta_\MESLE$ decreases to
\[
  \sigma^2 \left( \frac{\partial \hat \theta_\MESLE}{\partial \hat A} \right) \{ {\theta_{1:M}^{0:2}}^\top W^{adj} \theta_{1:M}^{0:2} + w_{M+1}^{adj} \theta_{M+1}^{0:2} {\theta_{M+1}^{0:2}}^\top \}^{-1} \left( \frac{\partial \hat \theta_\MESLE}{\partial \hat A} \right)^\top
\]
where $w_{M+1}^{adj}$ is the adjusted weight for the new simulation point.
We select the next simulation point that minimizes a scaled total variation of $\hat\theta_\MESLE$, defined as
\begin{equation}
  \mathrm{STV} := \mathrm{Tr}\left\{-\hat c^{-1} \left( \frac{\partial \hat \theta_\MESLE}{\partial \hat A} \right) \{ {\theta_{1:M}^{0:2}}^\top W^{adj} \theta_{1:M}^{0:2} + w_{M+1}^{adj} \theta_{M+1}^{0:2} {\theta_{M+1}^{0:2}}^\top \}^{-1} \left( \frac{\partial \hat \theta_\MESLE}{\partial \hat A} \right)^\top\right\}.
  \label{eqn:STV}
\end{equation}
Here $\mathrm{Tr}$ denotes the trace of a matrix and $-\hat c$ is the curvature of the fitted quadratic function before the addition of the new simulation point.
The scaled total variation (STV) of the parameter estimator incorporates inferential connections between parameter components and their relative scales by multiplying the approximated Monte Carlo variance by $-\hat c^{-1}$.
Algorithm~\ref{alg:optDesign} summarizes the automatic design point selection procedure.
The next simulation point, $\theta_{M+1}$, can be determined using standard numerical optimization routines such as the Broyden–Fletcher–Goldfarb–Shanno (BFGS) algorithm \cite{fletcher1987practical}.
Section~\ref{sec:supp_autotuning} of the supplementary text provides numerical demonstrations and additional details on this algorithm.
The method is implemented in the \texttt{optDesign} function of the \texttt{sbim} package.

\section{Discussion}\label{sec:discussion}

We developed a scalable inference method for models implicitly defined using simulators.
Implicitly defined, mechanistic models for the processes of interest can enable interpretable inference and accurate prediction.
However, exact simulation-based inference for these models is often impossible due to exponentially scaling Monte Carlo errors.
Our method employs a metamodel for the Monte Carlo log-likelihood estimator, enabling highly scalable inference.
We propose a principled approach to uncertainty quantification, which is often not addressed by recent methods using machine learning techniques that train surrogate models approximating the distribution of observations.
Our metamodel-based approach also differs from the synthetic likelihood method proposed by Wood \cite{wood2010statistical}, in that our metamodel describes the distribution of the log-likelihood estimator rather than that of summary statistics selected using domain knowledge.

We leave a few ideas that stem from the current work for future projects.
First, the assumptions in our metamodel may be relaxed to account for heteroskedasticity of log-likelihood estimators.
The extended model is likely to pose challenges in parameter estimation and uncertainty quantification, which might be addressed using suitable computational Bayesian methods.
Second, our inference framework using a metamodel for the log-likelihood estimator may be applied to likelihood estimates obtained from recently developed tools such as normalizing flows.
This application may benefit from a semiparametric extension of the metamodel, incorporating Gaussian-process-valued fluctuations around the quadratic approximation of the mean function.

The parameter inference methods developed in this paper are implemented in the \textsf{R} package \texttt{sbim} (\url{https://CRAN.R-project.org/package=sbim}).

\section*{Acknowledgments}
The author is grateful to Edward Ionides for his comments on early drafts of this manuscript.

\bibliographystyle{abbrv}
\bibliography{references}

\includepdf[pages=-]{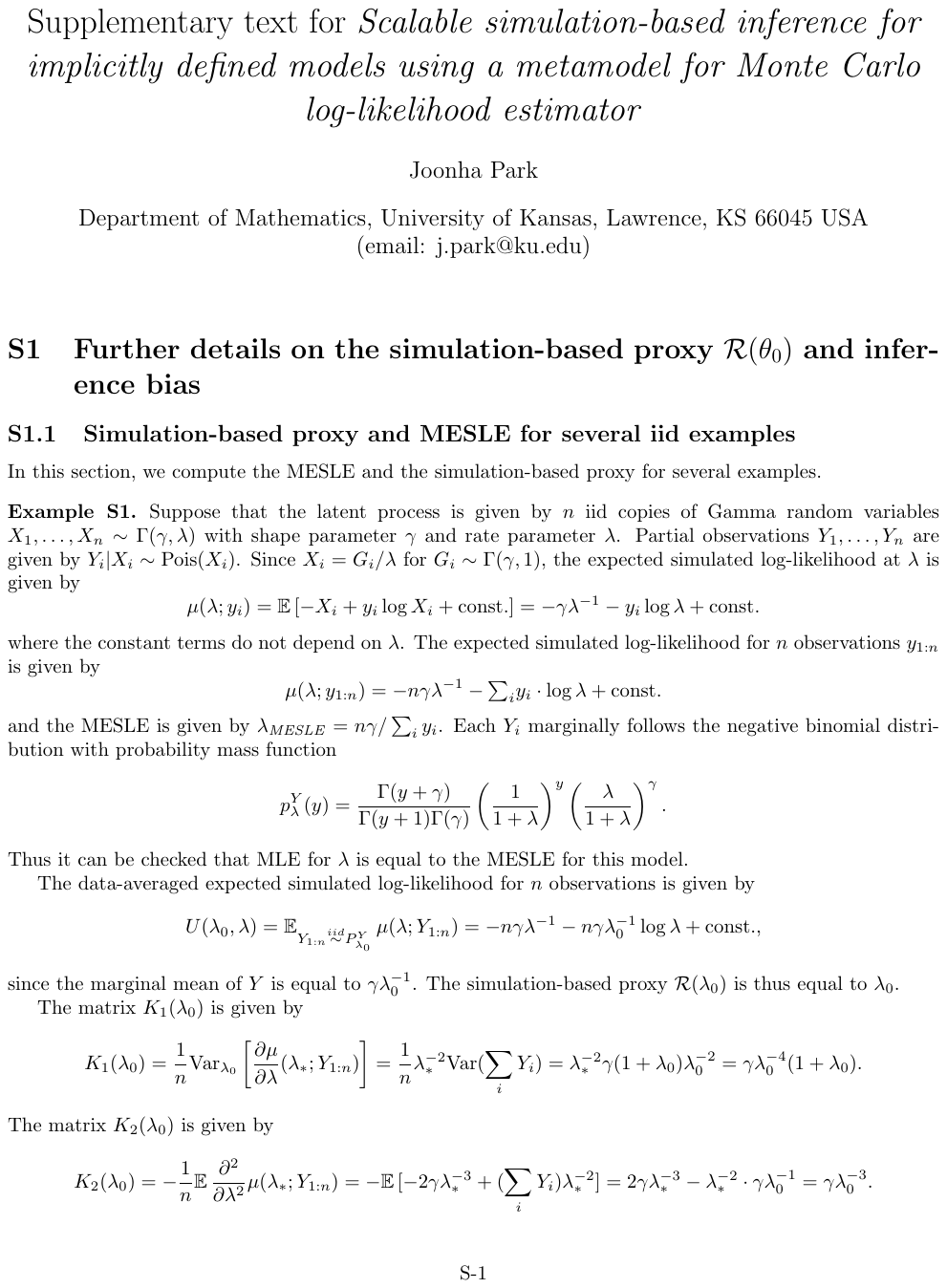}
\end{document}